\documentclass[11pt,fleqn,twoside]{article}
\usepackage{amsfonts,amssymb,latexsym}
\makeatletter
\newcommand{\prava}{\footnotesize\it
\begin{flushright}
\begin{minipage}{18cm}
Copyright \copyright 1998 by Boris. A. Kupershmidt
\end{minipage}
\end{flushright}}

\newcommand{\name}[1]{\begin{flushleft}
                       \LARGE \bf #1
                       \end{flushleft}\vspace{-3mm}}

\newcommand{\Author}[1]{\begin{flushleft}
                       \it #1 \end{flushleft}}

\newcommand{\Adress}[1]{\begin{flushleft}
                       \it #1 \end{flushleft}}

\newcommand{\Date}[1]{\begin{flushleft}
                      \small  \it #1 \end{flushleft}}

\newcommand{\ehkol}{Author \ name}
\newcommand{\ohkol}{Article \ name}
\renewcommand{\@evenhead}{
\hspace*{-3pt}\raisebox{-15pt}[\headheight][0pt]{\vbox{\hbox to \textwidth
{\thepage \hfil \ehkol}\vskip4pt \hrule}}}
\renewcommand{\@oddhead}{
\hspace*{-3pt}\raisebox{-15pt}[\headheight][0pt]{\vbox{\hbox to \textwidth
{\ohkol \hfil \thepage}\vskip4pt\hrule}}}
\renewcommand{\@evenfoot}{}
\renewcommand{\@oddfoot}{}

\newcommand{\be}{\begin{equation}}
\newcommand{\ee}{\end{equation}}
\newcommand{\ba}{\hspace*{-5pt}\begin{array}}
\newcommand{\ea}{\end{array}}

\newcommand{\ds}{\displaystyle}
\makeatother

\font\BoldMath=cmmib10 scaled \magstep1

\newcommand{\pmp}{\mbox{\BoldMath \char 112}}
\newcommand{\pmx}{\mbox{\BoldMath \char 120}}

\begin{document}
\setcounter{page}{245}

\thispagestyle{empty}

\renewcommand{\ehkol}{B.A. Kupershmidt}
\renewcommand{\ohkol}{Quantum Dif\/ferential Forms}

\begin{flushleft}
\footnotesize \sf
Journal of Nonlinear Mathematical Physics \qquad 1998, V.5, N~3,
\pageref{kupershmidt_1-fp}--\pageref{kupershmidt_1-lp}.
\hfill {\sc Article}
\end{flushleft}

\vspace{-5mm}

\renewcommand{\footnoterule}{}
{\renewcommand{\thefootnote}{}
\footnote{\prava}}

\name{Quantum Dif\/ferential Forms}\label{kupershmidt_1-fp}

\Author{Boris A. KUPERSHMIDT}

\Adress{The University of Tennessee Space Institute, Tullahoma, TN
37388  USA\\
E-mail: bkupersh@utsi.edu}

\Date{Received March 5, 1998}

\begin{flushright}
\begin{minipage}{7.3cm}
\small  \bfseries \itshape
Dedicated with gratitude to my teacher, Alexander Mikhailovich Vinogradov, on oc\-ca\-sion of his 60\,$^{th}$ anniversary.
\end{minipage}
\end{flushright}

\begin{abstract}
\noindent
Formalism of dif\/ferential forms is developed
for a variety of Quantum and noncommutative situations.\end{abstract}

\centerline{\bf Contents}

\begin{enumerate}

\item[\S~1.]  Introduction.

\item[\S~2.] Dif\/ferential forms over noncommutative polynomial rings.
\item[\S~3.] Noncommutative Lie derivatives.
\item[\S~4.] ${\bf Z}_2$--graded picture:  superdif\/ferential forms.
\item[\S~5.] $h$-Quantum spaces.
\item[\S~6.] Quantum Clebsch representations.
\item[\S~7.] Dif\/ferential forms on Lie algebras.
\item[\S~8.] The Lie algebra $af\!f(1)$ and its generalizations.
\item[\S~9.] The Lie algebra $gl(V)$.
\item[\S~10.] The Lie algebra $so(n)$.
\item[\S~11.] $Q$-Quantum spaces.
\item[\S~12.] $Q$-Quantum spaces and discrete groups.
\end{enumerate}

\subsection*{\S~1.  Introduction}
In its appearance, the algebraic apparatus of Quantum mechanics seems quite dissimilar from the familiar powerful machinery of Classical
mechanics/calculus of functions of several variables.
The crucial dif\/ference stems from the variables $p$'s and $q$'s
no longer commuting between themselves, thus rendering useless all
the comfortable tools of commutative mathematics.
Or so it seems, though it's mostly true.  But not entirely.
At any rate, the practical problems of Quantum mathematics,
for example those of Quantum integrable systems, require one
to establish missing Quantum analogs of versatile Classical tools.
This paper represents the second part of the project to develop such
tools; the f\/irst part [11] has dealt with motion equations.  Here
I take up the problem of constructing Quantum dif\/ferential forms,
the exterior dif\/ferential $d$, the Poincar\'e
Lemma, and various useful maps and relations between these.

As in the preceding paper, the basic philosophy is to look at
everything with noncommutative eyes and to utilize useful
noncommutative constructions whenever feasible.  The next two Sections
can be considered as a deleted Appendix from the noncommutative
textbook [12]; they set up the dif\/ferential forms, Lie derivatives,
and the Poincar\'e Lemma in general noncommutative polynomial rings.
Section~4 generalizes
all that to the ${\bf Z}_2$-graded case, and in the process establishes
what I think is the true form of the classical E.~Cartan formula for
the exterior  dif\/ferential $d$.

One of the main tools used in \S\S~2--4 is a construction of the homotopy operator.  Such an operator no longer exists in Quantum mechanics, \S~5; to establish there the Poincar\'e Lemma, I use instead elementary arguments of normal quantization.

\S~6 establishes a Quantum version of what is called {\it Clebsch representations} in [10],  -- but only for {\it finite-dimensional} Lie algebras, not dif\/ferential ones.  It's a bit unclear to me at the moment how to quantize the dif\/ferential case, or indeed if it is at all possible.  The device of Quantum Clebsch representations allows one to derive plausible rules for the generators and relations of a
dif\/ferential-forms complex attached to a f\/inite-dimensional Lie
algebra ${\cal{G}}$ with its f\/ixed representation on a vector space $V$; this is the subject of \S~7.  In contrast to the familiar complex of dif\/ferential forms associated to ${\cal{G}}$ and $V$, we get now a variety of Quantum-inspired ghosts.  For very special Lie algebras these ghosts can be avoided, as is done \S\S~8, 9 for the af\/f\/ine Lie algebra $af\!f(1)$ and the Lie algebra $gl(V)$ respectively; for the Lie algebra $so(V)$, the number of ghosts can be reduced, \S~10.

\S\S~11, 12 consider the Quantum spaces of $Q$-type, where the commutation relations between the variables $x_i$'s are of the form\[
x_i x_j = Q_{ij} x_j x_i, \qquad  \forall \ i, j,
\]with some invertible constants $Q_{ij}$'s.  These are the typical relations of Quantum vector spaces in the theory of Quantum Groups.  In \S~12 the variables $x_i$'s depend also on a discrete lattice index.  This prepares the grounds for the Quantum Variational calculus, the subject of a future paper.
\renewcommand{\thesection}{\arabic{section}}
\setcounter{section}{2}
\setcounter{equation}{0}
\renewcommand{\theequation}{\arabic{section}.\arabic{equation}}

\subsection*{\S~2.  Dif\/ferential forms over noncommutative polynomial rings}

Let $R$ be a f\/ixed associative ring with an unity and a ${\bf Q}$-algebra, -- the algebra of coef\/f\/i\-ci\-ents.  Denote by $R\langle x\rangle = R\langle
x_1,\ldots, x_n \rangle$ the ring of polynomials in the {\it noncommuting} variables $x_1,\ldots, x_n$; the coef\/f\/icients from $R$ {\it do commute} with the $x$'s.
The ring, and a $R\langle x\rangle $-bimodule, of dif\/ferential forms
on $R\langle x\rangle $, denoted $\Omega^* =  \Omega^* R\langle x\rangle$ is the noncommutative ring\be
\Omega^* R\langle x\rangle  = R\langle x,y\rangle  =
 R \langle x_1,\ldots, x_n, y_1,\ldots, y_n \rangle,
\ee$y_i$ denoting $dx_i$.  The dif\/ferential $d: R\langle x\rangle
\rightarrow \Omega^* R\langle x\rangle$
is an $R$-linear map and a derivation, satisfying the commutation rule\be
d x_i = y_i + x_i d, \qquad i = 1,\ldots,n.
\ee The wedge product $\mbox{sign}\;\wedge$ is suppressed from the
notation as not pertinent or advantageous.

There are various grading degrees attached to an element\[
\omega = \left\{\sum f_1 y_{i(1)} f_2 y_{i(2)} \ldots f_{\ell + 1} \ |
\ f_s \in R\langle x\rangle  \right\}
\]
from $\Omega^* R\langle x\rangle$.  Namely, the $x$-degree $p_x
(\omega)$, and the $dx$-degree $p_y(\omega)$.  Thus,
$\Omega^*R\langle x\rangle $ is bigraded,\be
\Omega^* = \oplus \Omega^{p,q},
\ee with\be
\Omega^{0,0} = R, \qquad  {\mathop{\oplus}_{p}} \Omega^{p,0} = R\langle x\rangle , \qquad {\mathop{\oplus}_{p}}
\Omega^{p,q} = : \Omega^q.
\ee
We next extend the dif\/ferential $d$ to act on the whole ring of dif\/ferential forms $\Omega^*$, by the commutation relations\setcounter{equation}{4}
\renewcommand{\theequation}{\arabic{section}.\arabic{equation}{\rm a}}
\be
dx_i = x_i d + y_i, \qquad  i = 1,\ldots, n,
\ee
\setcounter{equation}{4}
\renewcommand{\theequation}{\arabic{section}.\arabic{equation}{\rm b}}\be
dy_i = - y_i d, \qquad i = 1,\ldots, n,
\ee
\setcounter{equation}{4}
\renewcommand{\theequation}{\arabic{section}.\arabic{equation}{\rm c}}\be
dr = rd, \qquad d (r) = 0, \qquad \forall \ r \in R.
\ee Thus, $d$ becomes a graded derivation, of the bi-degree $(-1,1)$, satisfying the relation
\setcounter{equation}{5}
\renewcommand{\theequation}{\arabic{section}.\arabic{equation}}\be
d (\omega_1 \omega_2) = d (\omega_1) \omega_2 + (-1)^{p_{y }(\omega_{1})} \omega_1 d(\omega_2), \qquad \forall \ \omega_1, \omega_2 \in \Omega^*. \ee

\setcounter{equation}{7}

\noindent
{\bf Lemma 2.7.}  \be
d^2 = 0 \qquad {\rm on} \quad \Omega^* R\langle x\rangle.
\ee
\noindent
{\bf Proof.}  From formula (2.5) we f\/ind that\setcounter{equation}{8}
\renewcommand{\theequation}{\arabic{section}.\arabic{equation}{\rm a}}
\be
d^2 x_i = d\circ(x_i d + y_i) = (x_i d + y_i) d - y_i d = x_i d^2,
\ee \setcounter{equation}{8}
\renewcommand{\theequation}{\arabic{section}.\arabic{equation}{\rm b}}
\be
d^2 y_i = - dy_i d = y_i d^2,
\ee \setcounter{equation}{8}
\renewcommand{\theequation}{\arabic{section}.\arabic{equation}{\rm c}}
\be
d^2 r = rd^2, \qquad d^2 (r) = 0. \qquad  \mbox{\rule{2mm}{4mm}}
\ee 
\setcounter{equation}{9}
\renewcommand{\theequation}{\arabic{section}.\arabic{equation}}

We now shall examine whether every closed form $\omega, \ d(\omega) = 0$, is exact, $\omega = d (\nu)$ for some $\nu$.
Let us introduce a new variable $x_{n+1}$.  Call it $t$.  Let $t$
{\it commute} with everything.  Denote $dt$ by $\tau$.  Let $\tau$ also commute with everything, in the graded-dif\/ferential sense:\be
\tau \omega = (-1)^{p_{y}(\omega)} \omega \tau.
\ee To be a little bit less casual, let us adjoin $x_{n+1}$ and $\tau = y_{n+1}$ to $\Omega^* R\langle x\rangle $ without any assumptions of
commutatively apart from the def\/ining relations (2.5), and denote\setcounter{equation}{10}
\renewcommand{\theequation}{\arabic{section}.\arabic{equation}{\rm a}}
\be
a_i = tx_i - x_i t, \qquad  i = 1,\ldots, n,
\ee \setcounter{equation}{10}
\renewcommand{\theequation}{\arabic{section}.\arabic{equation}{\rm b}}
\be
b_\alpha = ty_\alpha - y_\alpha t, \qquad \alpha = 1,\ldots, n+ 1,
\ee \setcounter{equation}{10}
\renewcommand{\theequation}{\arabic{section}.\arabic{equation}{\rm c}}
\be
c_\alpha = x_\alpha \tau - \tau x_\alpha, \qquad \alpha = 1,\ldots, n + 1,
\ee \setcounter{equation}{10}
\renewcommand{\theequation}{\arabic{section}.\arabic{equation}{\rm d}}
\be
e_\alpha = \tau y_\alpha + y_\alpha \tau, \qquad \alpha = 1,\ldots, n+1.
\ee Then an easy check shows that\setcounter{equation}{11}
\renewcommand{\theequation}{\arabic{section}.\arabic{equation}{\rm a}}
\be
da_i = a_id + b_i - c_i,
\ee 
\setcounter{equation}{11}
\renewcommand{\theequation}{\arabic{section}.\arabic{equation}{\rm b}}\be
d b_\alpha = - b_\alpha d + e_\alpha,
\ee 
\setcounter{equation}{11}
\renewcommand{\theequation}{\arabic{section}.\arabic{equation}{\rm c}}\be
d c_\alpha = - c_\alpha d + e_\alpha,
\ee \setcounter{equation}{11}
\renewcommand{\theequation}{\arabic{section}.\arabic{equation}{\rm d}}
\be
e_\alpha d = d e_\alpha.
\ee Thus, we can indeed self-consistently allow $t$ and $\tau$ to commute with everything.
\setcounter{equation}{12}
\renewcommand{\theequation}{\arabic{section}.\arabic{equation}}

Next, formula (2.10) shows that (when characteristic $\not= 2$)\be
\tau^2 = 0.
\ee Thus,\be
R \langle x, t, y, \tau \rangle =
R \langle x, y\rangle [t] \ \oplus \ \tau R\langle x,y\rangle [t].
\ee 
In other words, every element $\omega$ of\be
{\overline \Omega\,}^* = R\langle x,t,y, \tau\rangle
\ee can be uniquely decomposed as\be
\omega = \omega_+ + \tau \omega_- , \qquad \omega_\pm \in \Omega^* [t]. \ee 

Now, let $I: {\overline \Omega\,}^* \rightarrow \Omega^*$ be the following $R$-linear map of $p_y$-degree  $-1$:\be
I (\omega) = \int\limits^1_0 dt \,\omega_-,
\ee where, for $\nu \in \Omega^*, $\be
\int\limits^1_0 dt \left(t^m \nu\right) = {1 \over m+1} \nu, \qquad   \forall \ m \in {\bf Z}_+, \ee 
The map $I$, as we shall see  presently, satisf\/ies all the properties of a homotopy operator (see , e.g., [3].)
Denote by $A_t: \Omega^* \rightarrow {\overline \Omega\,}^*$ the ring
homomorphism over $R$, def\/ined on the polynomial generators of
$\Omega^*$ by the rule:\setcounter{equation}{18}
\renewcommand{\theequation}{\arabic{section}.\arabic{equation}{\rm a}}
\be
A_t (x_i) = t x_i, \qquad i = 1,\ldots, n.
\ee 
\setcounter{equation}{18}
\renewcommand{\theequation}{\arabic{section}.\arabic{equation}{\rm b}}\be
A_t (y_i) = ty_i + \tau x_i, \qquad i = 1,\ldots, n.
\ee Thus, $A_t$ commutes with the operators $d$ in $\Omega^*$ and ${\overline \Omega\,}^*$: \setcounter{equation}{19}
\renewcommand{\theequation}{\arabic{section}.\arabic{equation}}
\be
d A_t = A_t d: \ \Omega^* \rightarrow {\overline \Omega\,}^*,
\ee because formulae (2.19) imply that\setcounter{equation}{20}
\renewcommand{\theequation}{\arabic{section}.\arabic{equation}{\rm a}}
\be
(d A_t - A_t d) x_i = t x_i (d A_t - A_t d),
\ee 
\setcounter{equation}{20}
\renewcommand{\theequation}{\arabic{section}.\arabic{equation}{\rm b}}\be
(d A_t - A_t d) y_i = - (ty_i + \tau x_i) (d A_t - A_t d).
\ee 
\setcounter{equation}{22}
\renewcommand{\theequation}{\arabic{section}.\arabic{equation}}

\noindent
{\bf Homotopy Formula 2.22.} {\it For any $\omega \in {\overline \Omega\,}^*, $\be
d I (\omega) + Id (\omega) = \omega_+|_{t = 1} - \omega_+|_{t = 0}. \ee 
}

\noindent
{\bf Proof.} By formula (2.16), it's enough to verify the homotopy formula (2.23) for two cases:
\setcounter{equation}{23}
\renewcommand{\theequation}{\arabic{section}.\arabic{equation}{\rm A}}\be
(A) \quad \omega = t^m \nu, \qquad m \in {\bf Z}_+, \quad \nu \in
\Omega^*;
\ee 
\setcounter{equation}{23}
\renewcommand{\theequation}{\arabic{section}.\arabic{equation}{\rm B}}\be
(B) \quad \omega = t^m \tau \nu, \qquad m \in {\bf Z}_+, \quad
\nu \in \Omega^*. \ee 

For the case ($A$), we have $\omega = \omega_+$, so that $I (\omega) = 0,$ and then
\setcounter{equation}{24}
\renewcommand{\theequation}{\arabic{section}.\arabic{equation}$\ell$}\be
Id (\omega) = I \left(t^m d\nu + mt^{m-1} \tau \nu\right)
= I \left(mt^{m-1} \tau \nu\right) = \int\limits^1_0 m t^{m-1} dt \nu
= \left(1 - \delta^0_m\right) \nu,
\ee while the $LHS$ of formula (2.23) yields\setcounter{equation}{24}
\renewcommand{\theequation}{\arabic{section}.\arabic{equation}$r$}
\be
t^m \nu|_{t=1} - t^m \nu|_{t = 0} = \nu \left(1 - \delta^0_m\right).
\ee 
{\samepage
For the case $(B)$, we have $\omega_+ = 0,$ and then \setcounter{equation}{25}
\renewcommand{\theequation}{\arabic{section}.\arabic{equation}{\rm a}}
\be
d I (\omega) = d \left(\int\limits^1_0 dt \,t^m \nu\right) =
d \left({1 \over m+1} \nu\right) = {1 \over m+1} d (\nu),
\ee \setcounter{equation}{25}
\renewcommand{\theequation}{\arabic{section}.\arabic{equation}{\rm b}}
\be
I (d \omega) = I \left(- t^m \tau d(\nu)\right) =
\int\limits^1_0 dt \; t^m d (\nu) = - {1 \over m+1} d(\nu),
\ee so that\setcounter{equation}{25}
\renewcommand{\theequation}{\arabic{section}.\arabic{equation}{\rm c}}
\be
(Id + dI) (\omega) = 0,
\ee while the $RHS$ of formula (2.23) vanishes because $\omega_+ = 0$.
\rule{2mm}{4mm}
}
\setcounter{equation}{27}
\renewcommand{\theequation}{\arabic{section}.\arabic{equation}}

\medskip

\noindent
{\bf Corollary 2.27.} {\it Suppose $\omega \in \Omega^*$ is a closed form.  Then there exists a form $\nu \in \Omega^*$ such that\be
(\omega - d (\nu)) \in R.
\ee 
In particular, every closed form of a positive homogeneous $p_y$-degree  is exact.}
\medskip

\noindent
{\bf Proof.}  Suppose $\omega \in \Omega^*$ is closed, $d(\omega) = 0$.  Then, $A_t (\omega)$ is also closed, in ${\overline \Omega\,}^*$, in view of formula (2.20).  The homotopy formula (2.23) then yields:\be
(A_t (\omega_+)) |_{t=1} - d I (A_t (\omega)) = (A_t (\omega ))_+ |_{t=0}.
\ee But, by formula (2.19), \be
(A_t (\omega))_+ |_{t=1} = \omega, \qquad  \forall \ \omega \in \Omega^*, \ee 
\be
(A_t (\omega))_+ |_{t=0} = pr^{0,0} (\omega), \qquad  \forall \ \omega \in \Omega^*,
\ee where $pr^{(0,0)} (\omega) $ is the $x,y$-independent part of $\omega$, its $R$-part.  Thus, \be
\omega = d I A_t (\omega) + pr^{0,0} (\omega).
\ee 

\noindent
{\bf Remark 2.33.} Everything so far proven remains true if we replace polynomials by formal power series, in any one the
combinations
\setcounter{equation}{33}
\renewcommand{\theequation}{\arabic{section}.\arabic{equation}{\rm a}}\be
R\langle \langle x\rangle \rangle \langle y\rangle ,
\ee 
\setcounter{equation}{33}
\renewcommand{\theequation}{\arabic{section}.\arabic{equation}{\rm b}}\be
R\langle x\rangle \langle \langle y\rangle \rangle ,
\ee 
\setcounter{equation}{33}
\renewcommand{\theequation}{\arabic{section}.\arabic{equation}{\rm c}} \be
R\langle \langle x,y\rangle \rangle .
\ee 
\setcounter{equation}{35}
\renewcommand{\theequation}{\arabic{section}.\arabic{equation}}

\noindent
{\bf Example 2.35.} Suppose $n = 1$ and \be
\omega_1 = y (1 - y)^{-1}, \qquad \omega_2 = (1 - y)^{-1}.
\ee Then both these forms are closed:\be
d (\omega_1) = d (\omega_2) = 0,
\ee and
\setcounter{equation}{37}
\renewcommand{\theequation}{\arabic{section}.\arabic{equation}{\rm a}}\be
\omega_1 = d \left(x (1 - y)^{-1}\right),
\ee 
\setcounter{equation}{37}
\renewcommand{\theequation}{\arabic{section}.\arabic{equation}{\rm b}}\be
\omega_2 = 1 + d \left(x (1-y)^{-1}\right).
\ee 
\noindent
{\bf Remark 2.39.}  The emphasis in this Section was on the homotopy operator as the crucial ingredient in establishing the Poincar\'e Lemma.  This is a very ef\/f\/icient route, and it will be followed in other Sections dealing with dif\/ferential forms, -- whenever possible.   It won't be {\it always} possible, as we shall see  in Section~5 devoted to Quantum Mechanics proper; we shall have to use other means there.
\medskip

\noindent
{\bf Remark 2.40.}  The dif\/ferential forms in this Section appear as
independent objects quite apart from their actions on vector f\/ields.
The main reason the latter have not been invited to partake in the
feast is that they ef\/fectively disappear in various Quantum
versions, especially in f\/ield theories, by virtue of not being able
to preserve the relevant Quantum commutation relations.  But
interestingly enough, in the universal totally
noncommutative framework of this Section, one can develop the
formalism of Lie derivatives rather close to the traditional commutative one.  This will be done in the next Section.

\medskip

\noindent
{\bf Remark 2.41.}  The reader will notice that everything in this Section holds true if the number of the $x$-generators, $n$, is inf\/inite.  The same observation applies also to all that follows.

\setcounter{section}{3}
\setcounter{equation}{0}
\renewcommand{\theequation}{\arabic{section}.\arabic{equation}}

\subsection*{\S~3.  Noncommutative Lie derivatives}

In the commutative picture, one has the following formulae relating dif\/ferential forms, vector f\/ields, and the dif\/ferential $d$: \be
X(\omega) = d (X \rfloor \omega) + X \rfloor d (\omega), \ee 
\be
X (\omega) (Z_1,\ldots, Z_\ell) = X (\omega (Z_1,\ldots, Z_\ell)) - \sum^\ell_{\alpha = 1} \omega (Z_1,\ldots, [X, Z_\alpha],\ldots, Z_\ell), \ee 
\be
X (f) = d(f) (X),
\ee \be
\ba{l}
\ds d (\omega) (Z_1,\ldots, Z_{\ell + 1}) =
\sum\limits^{\ell + 1}_{\alpha = 1} (-1)^{\alpha + 1} Z_\alpha (\omega (Z_1,\ldots,\hat Z_\alpha, \ldots,
Z_{\ell + 1} )\\[4mm]\ds \qquad + \sum\limits_{\alpha < \beta}
(-1)^{\alpha + \beta} \omega ([Z_\alpha, Z_\beta], Z_1,\ldots, \hat
Z_\alpha, \ldots, \hat Z_\beta, \ldots, Z_{\ell +
1}).
\ea
\ee Here $X$ and $Z_i$'s are vector f\/ields on a (smooth) manifold $M$, $\omega \in \wedge^\ell (M)$ is a dif\/ferential $\ell$-form on $M$, $f \in \wedge^0 (M)$ is a function on $M$, $d: \ \wedge^i (M) \rightarrow \wedge^{i+1} (M)$ is the (exterior) dif\/ferential, $X(\omega)$ is the Lie derivative of the form $\omega$ w.r.t. the vector f\/ield $X$, the hat $\hat{}$ over an argument indicates that it is missing, and $X \rfloor  \omega$ is the interior product:\be
(X \rfloor \omega) (Z_1,\ldots, Z_{\ell -1}) = \omega (X, Z_1,
\ldots, Z_{\ell -1}), \qquad \forall \ \omega \in \wedge ^\ell (M).
\ee 
In this Section we establish noncommutative analogs of these classical formulae.

We start with the ring $C = C_x = R \langle x \rangle = R \langle
x_1, \ldots, x_n \rangle $ of Section~2.  Denote by $\mbox{Der} (C)$ the Lie
algebra of derivations of $C$ over $R$: \be
X (fg) = X(f)g + f X(g), \qquad   \forall \ f, g \in C.
\ee 
Obviously, every element $X \in \mbox{Der} (C)$ is uniquely def\/ined by its (arbitrary) values on the generators of the ring $C$:\setcounter{equation}{6}
\renewcommand{\theequation}{\arabic{section}.\arabic{equation}{\rm a}}
\be
X (x_i) = X_i, \qquad X_i \in C, \qquad i = 1,\ldots, n,
\ee \setcounter{equation}{6}
\renewcommand{\theequation}{\arabic{section}.\arabic{equation}{\rm b}}
\be
X (r) = 0, \qquad  \forall \ r \in R.
\ee 

We shall f\/ind very useful the following device.  Instead of requiring $X$ to be on {\it apriori} derivation, we simply postulate how an additive map $X: C \rightarrow C$ commutes with the generators of $C$:\setcounter{equation}{7}
\renewcommand{\theequation}{\arabic{section}.\arabic{equation}{\rm a}}
\be
X x_i = x_i X + X_i, \qquad i = 1,\ldots, n,
\ee 
\setcounter{equation}{7}
\renewcommand{\theequation}{\arabic{section}.\arabic{equation}{\rm b}}\be
X r = r X, \qquad  \forall \ r \in R,
\ee \setcounter{equation}{7}
\renewcommand{\theequation}{\arabic{section}.\arabic{equation}{\rm c}}
\be
X (r) = 0, \qquad \forall \ r \in R.
\ee 
\setcounter{equation}{9}
\renewcommand{\theequation}{\arabic{section}.\arabic{equation}}

\noindent
{\bf Lemma 3.9.} {\it An additive map $X:  C \rightarrow C$ satisfying properties (3.8) is in fact a derivation of $C$.}
\medskip

\noindent
{\bf Proof.} We have to show that\be
X (fg) - X (f)g - f X(g)
\ee vanishes for all, $f, g \in C$.  Let us f\/ix $g$, and let $f$ vary.  Denote, temporarily,\be
\{X, f\} = X (fg) - X(f) g - fX(g).
\ee By formulae (3.8b,c), \be
\{ X, r \} = 0, \qquad \forall \ r \in R.
\ee Now,\be
\ba{l}
\ds \{ X, x_i f\} = X (x_i fg) - X (x_i f) g - x_i f X (g) \\[2mm]
\qquad \ds {\mathop{=}\limits^{\mbox{\scriptsize \rm [by (3.8a)]}}}\ (x_i X + X_i) (fg) - (x_i X + X_i) (f) \cdot g - x_i f X (g) = x_i \{X, f\}.
\ea
\ee Thus, induction on $\mbox{deg}_x (f)$ shows that $\{X, f\} = 0$. \qquad
\rule{2mm}{4mm}

The same device easily proves formula (2.6).  Fix $\omega_2$, and denote\[
\{\omega\} = d (\omega \omega_2) - d (\omega) \omega_2 - (-1)^p \omega d (\omega_2), \qquad   p = \mbox{deg}_y (\omega).
\]Then, by formula (2.5c),\[
\{r\} = 0, \qquad  \forall \ r \in R,
\]and then\[
\ba{l}
\ds \{x_i \omega\} = d (x_i \omega \omega_2) - d (x_i \omega) \omega_2 - (-1)^p x_i \omega d (\omega _2) \\[2mm]
\ds \qquad {\mathop{=}\limits^{\mbox{\scriptsize \rm [by  (2.5a)]}}}\  (x_i d + y_i) (\omega \omega_2) - ((x_i d + y_i) (\omega)) \omega_2 - (-1)^p x_i \omega d (\omega_2) = x_i \{ \omega \},
\ea
\]
\[\ba{l}
\ds \{y_i \omega \} = d (y_i \omega \omega_2) - d (y_i \omega) \omega_2 - (-1)^{p+1} y_i \omega d (\omega_2) \\[2mm]
\ds \qquad {\mathop{=}\limits^{\mbox{\scriptsize \rm [by (2.5b)]}}} \ - y_i d (\omega \omega_2) + y_i d (\omega) \omega_2 + (-1)^py_i \omega d (\omega_2) = - y_i \{\omega \}.
\ea
\]Thus, $\{\omega \}$ vanishes identically.

Given a derivation $X \in \mbox{Der}(C)$, we now extend its action from $C$ onto $\Omega^* = \Omega^* C = R \langle  x, y\rangle$, by adding to the commutation rules (3.8) the relations\be
Xy_i = y_i X + d (X (x_i)), \qquad i = 1,\ldots, n.
\ee 
\setcounter{equation}{15}
\renewcommand{\theequation}{\arabic{section}.\arabic{equation}}

\noindent
{\bf Lemma 3.15}.  {\it {\rm (i)}  $X$ is a derivation of the ring
$\Omega^*$; {\rm (ii)}  On $\Omega^*$, $X$ commutes with the differential $d$:\be
Xd = dX .
\ee }

\noindent
{\bf Proof.}  (i)  We proceed exactly as in the Proof of Lemma (3.9), taking $f$ and $g$ now not from the ring $C= R\langle x\rangle$ but from the ring  $\Omega^* = R\langle x,y\rangle$.  We need only to determine what
$\{X, y_i f\}$ is.  So, \be
\ba{l}
\ds \{X, y_i f\} = X (y_i f g) - X (y_i f) g - y_i f X (g) \\[2mm]
\ds \qquad {\mathop{=}\limits^{\mbox{\scriptsize \rm [by 3.14)]}}} \ (y_i X + d (X_i)) (fg) - (y_i X + d (X_i)) (f) \cdot g - y_i f X(g) = y_i \{X, f \};
\ea\!\!\!
\ee (ii)  To prove formula (3.16) we note that\be
(X d - dX) (r) = 0, \qquad \forall \ r \in R,
\ee and then verify the relations\setcounter{equation}{18}
\renewcommand{\theequation}{\arabic{section}.\arabic{equation}{\rm a}}
\be
(Xd - dX) x_i = x_i (Xd - dX),
\ee 
\setcounter{equation}{18}
\renewcommand{\theequation}{\arabic{section}.\arabic{equation}{\rm b}} \be
(Xd - dX) y_i = - y_i (Xd - dX),
\ee Indeed, \[
\ba{l}
\ds (Xd - dX) x_i \ {\mathop{=}\limits^{\mbox{\scriptsize \rm [by
(2.5a),  (3.8a)]}}} \ X (x_i d + y_i) - d (x_i X + X_i) \\[2mm]
\ds \qquad {\mathop{=}\limits^{\mbox{\scriptsize \rm [by (3.14)]}}}
\ (x_i X + X_i) d + y_i X + d (X_i) - (x_i d + y_i) X - d (X_i) - X_i d
\\[2mm]
\ds \qquad = x_i (Xd - dX),
\ea
\]\[
\ba{l}
 (Xd - dX) y_i = X (-1) y_i d - d (y_i X + d (X_i)) \\[2mm]
\ds \qquad =  - (y_i X + d (X_i)) d + y_i dX + d (X_i) d = - y_i (Xd - dX). \qquad \mbox{\rule{2mm}{4mm}}
\ea
\]

We next def\/ine the interior product, inductively:\setcounter{equation}{19}
\renewcommand{\theequation}{\arabic{section}.\arabic{equation}{\rm a}}
\be
X \rfloor \omega = 0 \quad  \mbox{if} \quad  p_y (\omega ) = 0,
\ee \setcounter{equation}{19}
\renewcommand{\theequation}{\arabic{section}.\arabic{equation}{\rm b}}
\be
X \rfloor \left(\sum_{is} f_{is} y_i g_{is}\right) =
\sum_{is} f_{is} X_i g_{is}, \qquad  f_{is}, g_{is} \in C = R\langle x\rangle ,
\ee 
\setcounter{equation}{19}
\renewcommand{\theequation}{\arabic{section}.\arabic{equation}{\rm c}} \be
X \rfloor x_i \omega = x_i (X \rfloor \omega), \qquad \omega \in \Omega^*, \quad i = 1,\ldots, n,
\ee \setcounter{equation}{19}
\renewcommand{\theequation}{\arabic{section}.\arabic{equation}{\rm d}}
\be
X \rfloor y_i \omega = X_i \omega - y_i (X \rfloor \omega), \qquad
\omega  \in \Omega^*, \quad i = 1,\ldots, n,
\ee \setcounter{equation}{19}
\renewcommand{\theequation}{\arabic{section}.\arabic{equation}{\rm e}}
\be
X \rfloor r \omega = r (X \rfloor \omega), \qquad \omega \in
\Omega^*, \quad r \in R.
\ee Notice that formulae (3.20c,d,e) agree  with (and, together with the relation (3.20a), imply) the formula (3.20b).
\setcounter{equation}{21}
\renewcommand{\theequation}{\arabic{section}.\arabic{equation}}

\medskip

\noindent
{\bf Lemma 3.21.} {\it For any $\omega_1, \omega_2 \in \Omega^*$, \be
X \rfloor \omega_1 \omega_2 = (X \rfloor \omega_1) \omega_2 + (-1)^{p_{y}(\omega_{1})} \omega_1 (X \rfloor \omega_2).
\ee }

\noindent
{\bf Proof.} Fix $\omega_2$, denote $p = p_y (\omega_1), $ and set
\[\{ X, \omega_1\} = X \rfloor \omega_1 \omega_2 - (X \rfloor \omega_1) \omega_2 - (-1)^p \omega_1 (X \rfloor \omega_2).
\]Then\[
 \{ X, x_i \omega_1\} = X \rfloor x_i \omega_1 \omega_2 - ( X \rfloor x_i \omega_1) \omega_2 - (-1)^p x_i \omega_1 (X \rfloor \omega_2) \  {\mathop{=}\limits^{\mbox{\scriptsize \rm [by (3.20c)]}}}
\  x_i \{X, \omega_1 \},
\]\[
\ba{l}
\ds \{X, y_i \omega_1\} = X \rfloor y_i \omega_1 \omega_2 - (X \rfloor y_i \omega_1) \omega_2 - (-1)^{p+1} y_i \omega_1 (X \rfloor \omega_2) \\[2mm]\ds \qquad {\mathop{=}\limits^{\mbox{\scriptsize \rm [by (3.20d)]}}} \  X_i \omega_1 \omega_2 - y_i (X \rfloor \omega_1 \omega_2) - (X_i \omega_1) \omega_2 + y_i (X \rfloor \omega_1) \omega_2 + (-1)^p y_i \omega_1 (X \rfloor \omega_2) \\[2mm]
\ds \qquad = - y_i \{ X, \omega_1 \}.
\ea\!
\]It remains to notice that, for any $r \in R$, \[
\{X, r \} = X \rfloor r \omega_2 - (X \rfloor r) \omega_2 - r (X \rfloor \omega_2) \ {\mathop{=}\limits^{\mbox{\scriptsize \rm [by
(3.20a,e)]}}} \  0 . \qquad
\mbox{\rule{2mm}{4mm}}
\]

We now have all the tools nee ded to state noncommutative analogs of the classical formulae (3.1)--(3.4).  First, formula (3.3):
\setcounter{equation}{23}

\medskip

\noindent
{\bf Lemma 3.23.} {\it For any $X \in \mbox{\rm Der} (C)$ and $f \in C$,\be
 X(f) = X \rfloor d (f).
\ee 
}
\noindent
{\bf Proof.} Set\[
\{X, f\} = X (f) - X \rfloor d (f).
\]Obviously,\[
\{ X, r\} = 0, \qquad  \forall \ r \in R.
\]Now,\[
\ba{l}
\ds \{X, x_i f\} = X (x_i f) - X \rfloor d (x_i f) = X_i f + x_i X (f) - X \rfloor (y_i f + x_i d (f)) \\[2mm]\ds \qquad = X_i f + x_i X (f) - X_i f - x_i (X \rfloor d (f)) = x_i
\{X, f\}. \qquad \mbox{\rule{2mm}{4mm}}
\ea
\]
Next comes formula (3.1):
\setcounter{equation}{25}

\medskip

\noindent
{\bf Lemma 3.25.} {\it  For any $X \in \mbox{\rm Der} (C)$ and $\omega \in \Omega^*$, \be
X (\omega) = d (X \rfloor \omega) + X \rfloor d (\omega).
\ee 
}

\noindent
{\bf Proof.} (A)  Set\[
\{X, \omega\} = X (\omega) - d (X \rfloor \omega) - X \rfloor d (\omega). \]
By Lemma 3.23 and formula (3.20a),\[
\{X, \omega\} = 0 \qquad  {\rm{when}} \quad p_y (\omega) = 0.
\]A direct check then shows that\[
\{ X, x_i \omega \} = x_i \{X, \omega\}, \qquad\{X, y_i \omega\} = y_i \{X, \omega \}.
\](B)  Alternatively, if $\omega = d(f)$,  $f \in C$, then formula (3.26) becomes\[
Xd (f) = d (X \rfloor d (f))
\](since $d^2 = 0)$, and this is true in view of formula (3.24), since
$Xd = dX $ by formula (3.16).  Now, one easily checks that
\[
\{X, \omega_1 \omega_2 \} = \{X, \omega_1\} \omega_2 + \omega_1 \{ X, \omega_2 \},
\]and this implies that $\{X, \omega\}$ vanishes identically, since
$C$ and $d(C)$ generate
the whole ring $\Omega^*$. \qquad \rule{2mm}{4mm}

Formula (3.2) is next, but it is a good time to take a skew-symmetric pause.  Noncommutative dif\/ferential forms dif\/fer from their commutative counterparts most clearly in not being skewsymmetric; after all, what is skewsymmetric about the expressions\[
(dx_1)^2, \qquad  \exp (dx_1).
\]and so on?  Interestingly enough, the skewsymmetry re-appears  when dif\/ferential forms are considered in their action on the (poly-) vector f\/ields:
\setcounter{equation}{27}

\medskip

\noindent
{\bf Lemma 3.27.} {\it  For any $Z_1, Z_2 \in \mbox{\rm Der} (C)$ and $\omega \in \Omega^*$, \be
Z_1 \rfloor Z_2 \rfloor \omega = - Z_2 \rfloor Z_1 \rfloor \omega.\ee 
}

\noindent
{\bf Proof.} Pick any two elements $\omega_1, \omega_2 \in \Omega^*$.
By formula (3.22), with $p = \mbox{deg}_y (\omega_1),
$
\be
\ba{l}
\ds
Z_1 \rfloor Z_2 \rfloor \omega_1 \omega_2 = Z_1 \rfloor
((Z_2 \rfloor \omega_1) \omega_2 + (-1)^p \omega_1 (Z_2 \rfloor \omega_2))
= (Z_1 \rfloor Z_2 \rfloor \omega_1) \omega_2
\\[2mm]
\ds
\qquad  -(-1)^p
(Z_2 \rfloor \omega_1) (Z_1 \rfloor \omega_2) + (-1)^p (Z_1 \rfloor
\omega_1) (Z_1 \rfloor \omega_2) +  \omega_1 (Z_1 \rfloor Z_2 \rfloor
\omega_2).
\ea
\ee
Thus,
\be
\ba{l}
\ds
Z_1 \rfloor Z_2 \rfloor \omega_1
\omega_2 + Z_2 \rfloor Z_1 \rfloor
\omega_1 \omega_2 \\[2mm]
\ds
 \qquad = (Z_1 \rfloor Z_2
\rfloor \omega_1+Z_2 \rfloor Z_1 \rfloor \omega_1) \omega_2 + \omega_1
(Z_1 \rfloor Z_2 \rfloor \omega_2 + Z_2 \rfloor Z_1 \rfloor \omega_2
).
\qquad \mbox{\rule{2mm}{4mm}}
\ea
\ee 
\setcounter{equation}{31}

\noindent
{\bf Corollary 3.31.} {\it For any $Z_1,\ldots, Z_\ell \in \mbox{\rm
Der} (C)$ and $\omega \in \Omega^*$, \[
Z_1 \rfloor Z_2 \rfloor \ldots. \rfloor Z_\ell \rfloor \omega
\]is totally skewsymmetric w.r.t. the $Z$'s:  for any permutation $\sigma \in S_\ell$, \be
Z_{\sigma (1)} \rfloor \ldots \rfloor Z_{\sigma (\ell)} \rfloor \omega = (-1)^{s g n (\sigma)} Z_1 \rfloor \ldots \rfloor Z_{\ell } \rfloor \omega, \qquad \forall \ \sigma \in S_\ell.
\ee 
}

\setcounter{equation}{33}

\noindent
{\bf Example 3.33.}  Denote by $f \partial_i$ the element of $\mbox{Der} (C)$ acting on the generators of $C$ by the rule\be
(f \partial_i) (x_j) = f \delta_{ij}, \qquad \forall \ f \in C,
\ee and write simply $f \partial$ and $y$ instead of $f \partial _1$ and $y_1$ when $n = 1$.  Then\be
f \partial_i \rfloor g \partial _j \rfloor y_i y_j = [f, g] \ (= fg -
gf),
\ee \be
(a \partial_i + b \partial_j) \rfloor (cd_i + d \partial_j) \rfloor
y_i y_j = cb - ad,
\ee 
\setcounter{equation}{36}
\renewcommand{\theequation}{\arabic{section}.\arabic{equation}{\rm a}} \be
f \partial \rfloor y = f,
\ee \setcounter{equation}{36}
\renewcommand{\theequation}{\arabic{section}.\arabic{equation}{\rm b}}
\be
f \partial \rfloor g \partial \rfloor y^2 = gf - fg,
\ee \setcounter{equation}{36}
\renewcommand{\theequation}{\arabic{section}.\arabic{equation}{\rm c}}
\be
f \partial \rfloor g \partial \rfloor h \partial \rfloor y^3 = h gf + gfh + fhg - hfg - fgh - ghf,
\ee 
\setcounter{equation}{36}
\renewcommand{\theequation}{\arabic{section}.\arabic{equation}{\rm d}} \be
\partial \rfloor y^{2 \ell} = 0, \qquad  \ell \in {\bf{Z}}_+,
\ee 
\setcounter{equation}{36}
\renewcommand{\theequation}{\arabic{section}.\arabic{equation}{\rm e}}
\be
\partial \rfloor y^{2 \ell +1} = y ^{2 \ell}, \qquad
 \ell \in {\bf{Z}}_+. \ee 

Formula (3.2) has the following noncommutative form:
\setcounter{equation}{38}
\renewcommand{\theequation}{\arabic{section}.\arabic{equation}}

\medskip

\noindent
{\bf Lemma 3.38.} {\it For any $X, Z_1,\ldots, Z_\ell \in
\mbox{\rm Der}(C)$  and $\omega \in \Omega^*$, \be
Z_\ell \rfloor \ldots \rfloor Z_1 \rfloor X (\omega)
= X (Z_\ell \rfloor\ldots \rfloor Z_1 \rfloor
\omega) - \sum^\ell_{\alpha = 1} Z_\ell \rfloor
\ldots  [X,  Z_\alpha] \rfloor\ldots
 \rfloor Z_1 \rfloor \omega.
\ee }

\setcounter{equation}{40}

\noindent
{\bf Remark 3.40.} Notice that, in contradistinction to the commutative case, the dif\/ferential form $\omega$ in formula (3.39) does not have to be a $\ell$-form.
\medskip

\noindent
{\bf Proof.}  We f\/irst establish formula (3.39) for the case $\ell = 1$:
\be Z \rfloor X (\omega) = X (Z \rfloor \omega) - [X, Z] \rfloor \omega. \ee 
We shall prove formula (3.41) in 3 stages:

1)  The formula is obvious when $\mbox{deg}_y (\omega) = 0; $

2)  If $\omega = y_i$ then\[
Z \rfloor X (y_i) = Z \rfloor d (X_i) = Z (X_i),
\]while\[
X (Z \rfloor y_i) - [X, Z] \rfloor y_i = X (Z_i) - [X, Z]_i = X (Z_i) - (X (Z_i) - Z (X_i)) = Z (X_i);
\]
3)  Since $\Omega^*$ is generated by $C$ and the $y_i$'s, it's enough to check that if formula (3.41) holds for $\omega_1, \omega_2 \in \Omega^*$ then it also holds for $\omega = \omega_1 \omega_2$.  Denoting $p = \mbox{deg}_y
(\omega_1)$, we f\/ind\setcounter{equation}{41}
\renewcommand{\theequation}{\arabic{section}.\arabic{equation}$\ell$}
\be
\ba{l}
\ds Z \rfloor X (\omega_1 \omega_2) = Z \rfloor (X (\omega_1) \omega_2 + \omega_1 X (\omega_2)) = (Z \rfloor X (\omega_1)) \omega_2
\\[2mm]
\ds \qquad + (-1)^p X (\omega_1)
(Z \rfloor \omega_2) + (Z \rfloor \omega_1) X( \omega_2) + (-1)^p
\omega_1 (Z \rfloor X (\omega_2)),
\ea
\ee 
\setcounter{equation}{41}
\renewcommand{\theequation}{\arabic{section}.\arabic{equation}$r$} \be
\ba{l}
\ds X (Z \rfloor \omega_1 \omega_2) = X ((Z \rfloor \omega_1) \omega_2 + (-1)^p \omega_1 (Z \rfloor \omega_2)) = X (Z \rfloor \omega_1)
\omega_2 \\[2mm]\ds \qquad + (Z \rfloor \omega_1) X (\omega_2) + (-1)^p X (\omega_1)
(Z \rfloor \omega_2) + (-1)^p \omega_1 X (Z \rfloor \omega_2)
\\[2mm]
\ds \qquad - [X, Z] \rfloor \omega_1 \omega_2 = - ([ X, Z] \rfloor
\omega_1) \omega_2 - (-1)^p \omega_1 ([X, Z] \rfloor \omega_2).
\ea
\ee 

\setcounter{equation}{42}
\renewcommand{\theequation}{\arabic{section}.\arabic{equation}}

{\samepage
Adding all up, we get:\be
\ba{l}
Z \rfloor X (\omega_1 \omega_2) - X (Z \rfloor \omega_1 \omega_2) - [X, Z] \rfloor \omega_1 \omega_2 \\[2mm]\ds \qquad = (Z \rfloor X( \omega_1) - X (Z \rfloor \omega_1) - [X,
Z] \rfloor \omega_1) \omega_2 \\[2mm]
\ds \qquad + (-1)^p \omega_1 (Z \rfloor X (\omega_2) - X (Z \rfloor
\omega_2) - [X, Z] \rfloor \omega_2),
\ea
\ee as desired.}

With formula (3.41) behind us, we could take two routes to the general formula (3.39).  The longer route splits $x_i$ and $y_i$ from the left of $\omega $ and uses the formula\be
\hspace*{-8.5pt}
Z_1 \rfloor \ldots \rfloor Z_\ell \rfloor y_i \omega = (-1)^\ell y_i (Z_1 \rfloor \ldots \rfloor Z_\ell \rfloor \omega) +  \sum^\ell_{\alpha = 1} (-1)^{\ell - \alpha} y_i (Z_\alpha) (Z_1 \rfloor \ldots\hat Z_\alpha \ldots Z_\ell \rfloor \omega).
\ee The shorter route uses induction on $\ell$: For $\ell = 1$, formula (3.39) turns into already proven formula (3.41), and then\[\hspace*{-26.5pt}
\ba{l}
\ds Z_{\ell +1} \rfloor \ldots \rfloor Z_1  \rfloor X (\omega)
\ {\mathop{=}\limits^{\mbox{\scriptsize  \rm [by (3.39)]}}} \ Z_{\ell +1} \rfloor \left\{ X (Z_\ell \rfloor \ldots \rfloor Z_1 \rfloor \omega) - \sum\limits^\ell_{\alpha = 1} Z_\ell \rfloor \ldots \rfloor
[X, Z_\alpha ] \rfloor \ldots
 \rfloor Z_1 \rfloor \omega \right\} \\[3mm]
\ds {\mathop{=}\limits^{\mbox{\scriptsize \rm [by 3.41)]}}} \ X (Z_{\ell +1} \rfloor\ldots \rfloor Z_1 \rfloor \omega) - [X, Z_{\ell + 1} ] \rfloor Z_{\ell} \rfloor \ldots \rfloor Z_1 \omega - \sum\limits^\ell _{\alpha =1} Z_{\ell +1} \rfloor Z_\ell \rfloor
\ldots \rfloor [X, Z_\alpha ] \rfloor \ldots \rfloor Z_1 \rfloor \omega \\[3mm]\ds \qquad  = X (Z_{\ell + 1} \rfloor \ldots \rfloor Z_1 \rfloor \omega) - \sum\limits^{\ell = 1}_{\alpha =1} Z_{\ell + 1}
\rfloor \ldots \rfloor [X, Z_\alpha] \rfloor \ldots\rfloor Z_1 \rfloor \omega,
\ea
\]which is formula (3.39) with $\ell$ replaced by $\ell + 1$. \qquad
\rule{2mm}{4mm}

\medskip

We are now ready to handle the last of the classical formulae
(3.1)--(3.4), E.~Cartan's formula (3.4).  We start with formula
(3.26) rewritten in the form \be
Z_1 \rfloor d (\omega) = Z_1 (\omega) - d (Z_1 \rfloor \omega). \ee Applying the operation $Z_2 \rfloor $ to each side of formula (3.45),
we f\/ind \[
\ba{l}
\ds Z_2 \rfloor Z_1 \rfloor d (\omega) = Z_2 \rfloor Z_1 (\omega) - Z_2 \rfloor d (Z_1 \rfloor \omega) \\[2mm]
\ds \qquad {\mathop{=}\limits^{\mbox{\scriptsize \rm [by (3.41, 3.45)]}}} \ Z_1 (Z_2 \rfloor \omega) - [Z_1, Z_2 ] \rfloor \omega - Z_2 (Z_1 \rfloor \omega) + d (Z_2 \rfloor Z_1 \rfloor \omega) .
\ea
\]Thus,\be
Z_2 \rfloor Z_1 \rfloor d (\omega) = Z_1 (Z_2 \rfloor \omega) - Z_2 (Z_1 \rfloor \omega) + d (Z_2 \rfloor Z_1 \rfloor \omega) - [Z_1, Z_2] \rfloor \omega.
\ee 
We see  that in each of formulae (3.45), (3.46) we get an extra
$d$-term compared to the classical formula, -- because we have not
required that the $d$-degree  of $\omega$ be equal to the number of vector f\/ields $Z_i$'s.
\setcounter{equation}{47}

\medskip

\noindent
{\bf Lemma 3.47.} {\it For any $Z_1,\ldots,Z_\ell \in \mbox{\rm Der}
(C)$ and $\omega \in \Omega^*$,\be
\ba{l}
\ds Z_\ell \rfloor \ldots \rfloor Z_1 \rfloor d (\omega) =
\sum\limits^\ell_{\alpha = 1} (-1)^{\alpha + 1} Z_\alpha (Z_\ell \rfloor \ldots
\hat Z_\alpha \ldots \rfloor Z_1 \rfloor \omega) + (-1)^\ell d (Z_{\ell} \rfloor \ldots
 \rfloor Z_1 \rfloor \omega) \\[3mm]\ds \qquad + \sum\limits_{\alpha < \beta}
(-1)^{\alpha + \beta} (Z_\ell \rfloor \ldots \hat Z_{\beta} \ldots \hat Z_\alpha \ldots [Z_\alpha , Z_\beta] \rfloor
\omega).
\ea\hspace{-4.2pt}
\ee }

\noindent
{\bf Proof.} We use induction on $\ell$, the cases $\ell = 1,2$
having been verif\/ied by formulae (3.45) and (3.46) respectively.
Applying the operation $Z_{\ell +1} \rfloor$ to each side of formula (3.48), we f\/ind: \[
Z_{\ell + 1} \rfloor \ldots \rfloor Z_1 \rfloor d (\omega)
\]
\setcounter{equation}{48}
\renewcommand{\theequation}{\arabic{section}.\arabic{equation}{\rm a}} \be
\qquad = \sum^\ell_{\alpha =1} (-1)^{\alpha + 1} Z_{\ell + 1} \rfloor
Z_\alpha (Z_{\ell} \rfloor \ldots \hat Z_{\alpha}
\ldots Z_1 \rfloor \omega) \ee 
\setcounter{equation}{48}
\renewcommand{\theequation}{\arabic{section}.\arabic{equation}{\rm b}}
\be
\qquad + (-1)^\ell Z_{\ell +1} \rfloor d (Z_{\ell} \rfloor \ldots Z_1 \rfloor \omega)
\ee \setcounter{equation}{48}
\renewcommand{\theequation}{\arabic{section}.\arabic{equation}{\rm c}}
\be
\qquad + \sum_{\alpha < \beta \leq \ell} (-1)^{\alpha + \beta}
(Z_{\ell + 1} \rfloor \ldots \hat Z_{\beta} \ldots
\hat Z_\alpha \ldots [Z_\alpha, Z_\beta] \rfloor \omega).
\ee By formula (3.41), the sum (3.49a) can be transformed as
\setcounter{equation}{48}
\renewcommand{\theequation}{\arabic{section}.\arabic{equation}{\rm a1}} \be
\sum^\ell_{\alpha = 1} (-1)^{\alpha +1} Z_\alpha (Z_{\ell + 1} \rfloor \ldots \hat Z_\alpha \ldots Z_1 \rfloor \omega)
\ee 
\setcounter{equation}{48}
\renewcommand{\theequation}{\arabic{section}.\arabic{equation}{\rm a2}}
\be
\qquad\qquad + \sum^\ell_{\alpha = 1} (-1)^\alpha ([Z_\alpha, Z_{\ell +1}] \rfloor Z_\ell \rfloor \ldots \hat Z_\alpha \ldots Z_1 \rfloor \omega).
\ee 

By formula (3.45), the second sum (3.49b) becomes\setcounter{equation}{48}
\renewcommand{\theequation}{\arabic{section}.\arabic{equation}{\rm b1}}
\be
(-1)^\ell Z_{\ell +1} (Z_\ell \rfloor \ldots \rfloor Z_1 \rfloor
\omega)
\ee \setcounter{equation}{48}
\renewcommand{\theequation}{\arabic{section}.\arabic{equation}{\rm b2}}
\be
\qquad + (-1)^{\ell +1} d (Z_{\ell + 1} \rfloor
\ldots \rfloor Z_1 \rfloor \omega ).
\ee Combining the terms in formulae (3.49a1), (3.49b1), we get
\setcounter{equation}{49}
\renewcommand{\theequation}{\arabic{section}.\arabic{equation}} \be
\sum^{\ell + 1}_{\alpha = 1} (-1)^{\alpha + 1} Z_\alpha (Z_{\ell + 1}  \rfloor \ldots \hat Z_\alpha \ldots
 \rfloor Z_1 \rfloor \omega).
\ee Rewriting the sum (3.49a2) as\be
\sum^\ell_{\alpha = 1} (-1)^{\alpha + \ell + 1} (Z_\ell \rfloor \ldots \hat Z_\alpha \ldots [Z_\alpha, Z_{\ell + 1} ] \rfloor\omega)
\ee and combining it with the sum (3.49c), we obtain\be
+ \sum_{\alpha < \beta \leq \ell + 1} (-1)^{\alpha + \beta} (Z_{\ell + 1} \rfloor \ldots \hat Z_\beta \ldots \hat Z_\alpha \ldots [Z_\alpha,
Z_\beta] \rfloor \omega).
\ee Adding up formulae (3.50), (3.49b2), (3.52), we recover the RHS of
formula (3.48) with $\ell$ replaced by $\ell + 1$. \qquad
\rule{2mm}{4mm}

\medskip

So far, we have paid no attention to many related subjects lurking in the shadows.  No mention has been made of homology (see , e.g.,
[5])
of the Lie algebra $\mbox{Der}(C)$ (or, more generally, $\mbox{Der}(\Omega^*, d)$,
see  below.) But one shouldn't ignore the ${\bf Z}_2$-graded nature of the ring $\Omega^*$:\be
\Omega^* = \Omega^*_e + \Omega^*_o.
\ee where $\omega \in \Omega^*$ is even or odd depending upon $p (\omega) = p_y (\omega) \ \mbox{mod}\; 2$ being respectively 0 or 1 in ${\bf Z}_2$.
Consequently, additive maps from $\Omega^*$ to $\Omega^*$ are also ${\bf Z}_2$-graded, and one can talk about ${\bf Z}_2$-graded derivations $Y \in \mbox{Der}
(\Omega^*)$:
\be
Y(\omega_1 \omega_2) = Y (\omega_1) \omega_2 + (-1)^{p (Y)p (\omega_{1})} \omega_1 Y (\omega_2), \qquad \forall \ \omega_1, \omega_2 \in
\Omega^*. \ee 
Since we already have the dif\/ferential $d$ acting on $\omega^*$ (as an {\it old} derivation, see  formula (2.6)), the most important subsuperalgebra in the Lie superalgebra $\mbox{Der} (\Omega^*)$ is\be
\mbox{Der} (\Omega^*, d) = \{ Y \in \;\mbox{Der} (\Omega^*) | \quad
 Y d = (-1)^{p(Y)} d Y\}. \ee 
For example,\be
d \in \mbox{Der} (\omega^*, d),
\ee and of course\be
\mbox{Der} (C) \subset \mbox{Der} (\Omega^*, d).
\ee $\mbox{Der} (C)$ is an even subsuperalgebra in $\mbox{Der} (\Omega^*,
d)$, but it is by no means {\it all} of the even part of $\mbox{Der} (\Omega^*, d)$.  All noncommutative formulae proved in this Section for elements $Z_i \in
\mbox{Der} (C)$ remain true for {\it even} elements $Z_i \in
\mbox{Der}
(\Omega^*,d)_e$, although this is not immediately obvious in view of the
commutators $[Z_i, Z_j]$ entering our formulae in places.  But we can
do better still, and consider the vector f\/ield arguments $X$ and
$Z_i$'s of {\it arbitrary} ${\bf Z}_2$-grading, whether even or odd.
On the second thought, we
could have started with the generators $x_i$'s of prescribed arbitrary
${\bf Z}_2$-grading $p(i)$.  And on the third thought, we could have
taken the coef\/f\/icient ring $R$ being ${\bf Z}_2$-graded as well.
This program is realized in the next Section.
\setcounter{section}{4}
\setcounter{equation}{0}
\renewcommand{\theequation}{\arabic{section}.\arabic{equation}}

\subsection*{\S~4. ${\bf Z}_2$-graded picture:  superdif\/ferential
forms}

Recall a few basic facts about superobjects.  Suppose $R$ and ${\cal
R}$ are ${\bf Z}_2$-graded associative rings, with ${\cal R}$ being an
$R$-algebra.  The latter means that\be
r \rho = (-1) ^{p(r)p(\rho)} \rho r, \qquad r \in R, \quad \rho \in
{\cal R}, \ee 
where $p(\cdot)$ is the ${\bf Z}_2$-degree  of $(\cdot)$.  A (left)
{\it derivation} of ${\cal R}$ over $R$ is an additive map $Z: {\cal R}
\rightarrow {\cal R}$ satisfying  the properties\setcounter{equation}{1}
\renewcommand{\theequation}{\arabic{section}.\arabic{equation}{\rm a}}
\be
Z (\rho_1 \rho_2) = Z (\rho_1) \rho_2 + (-1)^{p (Z) p ( \rho_1)}\rho_1 Z (\rho_2), \qquad \rho_1, \rho_2  \in {\cal R},
\ee \setcounter{equation}{1}
\renewcommand{\theequation}{\arabic{section}.\arabic{equation}{\rm b}}
\be
Z (r \rho) = (-1) ^{p (Z) p (r)} r Z (\rho), \qquad r \in {\cal R},  \quad \rho \in {\cal R},
\ee 
\setcounter{equation}{1}
\renewcommand{\theequation}{\arabic{section}.\arabic{equation}{\rm c}}\be
Z(r) = 0, \qquad  r \in R.
\ee Property (4.2c) assumes that $R$ has a unit element.  The set of all such derivations is denoted $\mbox{Der} ({\cal R}) = \mbox{Der}({\cal
R}/R)$.  It is a Lie superalgebra: if $Z_1, Z_2, Z_3 \in \mbox{Der} ({\cal R}$) then \setcounter{equation}{1}
\renewcommand{\theequation}{\arabic{section}.\arabic{equation}}
\be
[Z_1, Z_2] : =  Z_1 Z_2 - (-1)^{p(Z_1) p (Z_1)} Z_2 Z_1 = - (-1)^{p(Z_1)p Z_2)} [Z_2, Z_1]
\ee is also an element of $\mbox{Der}({\cal R}$), and \be
[Z_1, [Z_2, Z_3]] = [[Z_1, Z_2], Z_3] + (-1)^{p (Z_1)  p (Z_2)} [Z_2, [Z_1, Z_3]].
\ee The reader will notice the convention employed in ${\bf Z}_2$-graded formulae:  they are often written for ${\bf Z}_2$-homogeneous elements only.

We now take ${\cal R}= R\langle x\rangle = R\langle x_1,\ldots,
x_n\rangle $, with the $x_i$'s having arbitrarily prescribed ${\bf Z}_2$-gradings $p(i)$: \be
p (x_i) = p (i), \qquad i = 1,\ldots, n.
\ee The dif\/ferential $d: {\cal R} \rightarrow \Omega^* = R\langle
x,y\rangle $ is now def\/ined  as an odd map satisfying the properties\setcounter{equation}{5}
\renewcommand{\theequation}{\arabic{section}.\arabic{equation}{\rm a}}
\be
dx_i = (-1)^{p(i)} x_i d + y_i, \qquad i = 1,\ldots,n,
\ee \setcounter{equation}{5}
\renewcommand{\theequation}{\arabic{section}.\arabic{equation}{\rm b}}
\be
dr = (-1)^{p(r)} rd, \qquad r \in R,
\ee \setcounter{equation}{5}
\renewcommand{\theequation}{\arabic{section}.\arabic{equation}{\rm c}}
\be
d(r)= 0, \qquad r \in R.
\ee 
The generators $y_i$'s of $\Omega^*$ have the natural ${\bf Z}_2$-grading opposite to that of the $x_i$'s:\[
p (y_i) = p (x_i) + \overline 1 = p(i) + 1
\](we write 1 instead of $\overline 1$ in ${\bf Z}_2)$.  Extending the action of $d$ from ${\cal R}$ onto $\Omega^*$ we add to formulae (4.6) another one:\setcounter{equation}{5}
\renewcommand{\theequation}{\arabic{section}.\arabic{equation}{\rm d}}
\be
dy_i = - (-1)^{p(i)} y_i d, \qquad i = 1,\ldots, n.
\ee These relations imply that $d: \Omega^* \rightarrow \Omega^*$ is an odd derivation:\setcounter{equation}{6}
\renewcommand{\theequation}{\arabic{section}.\arabic{equation}}
\be
d (\omega_1 \omega_2) = d(\omega_1) \omega_2 + (-1)^{p (\omega_1)} \omega_1 d (\omega_2), \qquad \omega_1, \omega_2 \in \Omega^*,
\ee and that $d^2=0$ on $\Omega^*$.  (This and other easily checked facts in this Section are left to the reader.)

Since $\Omega^*$ is also an $R$-algebra, we have two Lie superalgebras: $\mbox{Der}({\cal R})$ and $\mbox{Der}(\Omega^*)$.  The latter is too
big, and we need only a part of it:\be
\mbox{Der} (\Omega^*, d) = \{ Z \in \mbox{Der} (\Omega^*) | \quad Zd
= (-1)^{p(Z)} Zd \}; \ee 
alternatively, we can describe such $Z$'s as additive maps $\Omega^* \rightarrow \Omega^*$ satisfying the relations\setcounter{equation}{8}
\renewcommand{\theequation}{\arabic{section}.\arabic{equation}{\rm a}}
\be
Z x_i = (-1)^{p (Z) p(i)} x_i Z + Z_i, \qquad Z_i = Z (x_i) \in
\Omega^*, \quad i = 1,\ldots, n,
\ee 
\setcounter{equation}{8}
\renewcommand{\theequation}{\arabic{section}.\arabic{equation}{\rm b}}\be
Z y_i = (-1)^{p(Z)} ( (-1)^{p (Z)p(i)} y_i Z + d (Z_i)), \qquad
i = 1,\ldots,  n,
\ee \setcounter{equation}{8}
\renewcommand{\theequation}{\arabic{section}.\arabic{equation}{\rm c}}
\be
Z r = (-1)^{p(Z)p(r)} r Z, \qquad r \in R,
\ee 
\setcounter{equation}{8}
\renewcommand{\theequation}{\arabic{section}.\arabic{equation}{\rm d}}\be
Z (r) = 0, \qquad r \in R.
\ee 
Let us f\/irst dispose of the Poincar\'e Lemma.  As in \S~2, we adjoin an {\it even} variable $x_{n+1} = t$ and let it commute with everything; its dif\/ferential $dt = \tau$ we also let (super) commute with everything:\setcounter{equation}{9}
\renewcommand{\theequation}{\arabic{section}.\arabic{equation}}
\be
\tau \omega = (-1)^{p(\omega)} \omega \tau, \qquad \omega \in {\overline \Omega\,}^*.
\ee Using again the unique decomposition\be
\omega = \omega_+ + \tau \omega_-, \qquad \omega \in {\overline \Omega\,}^* = R\langle  x,t,y,z\rangle , \qquad  \omega_\pm \in \Omega^* [t],
\ee we set\be
I (\omega) = \int\limits^1_0 dt \,\omega_- ,
\ee and def\/ine the even ring homomorphism $A_t : \Omega^* \rightarrow {\overline \Omega\,}^*$ by the rules\setcounter{equation}{12}
\renewcommand{\theequation}{\arabic{section}.\arabic{equation}{\rm a}}
\be
A_t (x_i) = tx_i, \qquad i = 1,\ldots, n,
\ee 
\setcounter{equation}{12}
\renewcommand{\theequation}{\arabic{section}.\arabic{equation}{\rm b}}\be
A_t (y_i) = ty_i + \tau x_i, \qquad i = 1,\ldots, n,
\ee \setcounter{equation}{12}
\renewcommand{\theequation}{\arabic{section}.\arabic{equation}{\rm c}}
\be
A_t (r) = r, \qquad r \in R.
\ee These rules imply that
\setcounter{equation}{13}
\renewcommand{\theequation}{\arabic{section}.\arabic{equation}{\rm a}}\be
(dA_t - A_t d) r = (-1)^{p (r)} r (dA_t - A_t d), \qquad r \in R, \ee 
\setcounter{equation}{13}
\renewcommand{\theequation}{\arabic{section}.\arabic{equation}{\rm b}}
\be
(dA_t - A_t d) x_i = (-1)^{p(i)} tx_i (dA_t - A_t d), \qquad
i = 1, \ldots, n, \ee 
\setcounter{equation}{13}
\renewcommand{\theequation}{\arabic{section}.\arabic{equation}{\rm c}}
\be
(dA_t - A_t d) y_i = (-1)^{p(i)+1} (ty_i + \tau x_i) (dA_t - A_t d),
\qquad i = 1,\ldots, n,
\ee and thus\setcounter{equation}{14}
\renewcommand{\theequation}{\arabic{section}.\arabic{equation}}
\be
A_t d = d A_t : \ \Omega^* \rightarrow {\overline \Omega\,}^*.
\ee 
The homotopy formula (2.23):\be
dI (\omega) + Id (\omega) = \omega_+|_{t=1} - \omega_+|_{t =0},
\qquad \forall \ \omega \in {\overline \Omega\,}^*,
\ee holds true with the same Proof as in \S~2.  Therefore, again as in \S~2,\be
d (\omega) = 0 \ \Rightarrow \ \omega = d I A_t (\omega) + pr^{(0,0)} (\omega), \qquad  \omega \in \Omega^*.
\ee 
Let us now turn to the Lie derivative formulae.  First, we def\/ine the operation $X \rfloor$, for $X \in \mbox{Der}(\Omega^*, d)$, by the rules\setcounter{equation}{17}
\renewcommand{\theequation}{\arabic{section}.\arabic{equation}{\rm a}}
\be
X \rfloor \omega = 0, \qquad p_y (\omega) = 0,
\ee 
\setcounter{equation}{17}
\renewcommand{\theequation}{\arabic{section}.\arabic{equation}{\rm b}}\be
X \rfloor r \omega = (-1)^{p (r) (p(X)+1)} r (X \rfloor \omega),
\qquad r \in R,
\ee \setcounter{equation}{17}
\renewcommand{\theequation}{\arabic{section}.\arabic{equation}{\rm c}}
\be
X \rfloor x_i \omega = (-1)^{p(i) (p (X)+1)} x_i (X \rfloor \omega), \qquad i = 1,\ldots, n,
\ee \setcounter{equation}{17}
\renewcommand{\theequation}{\arabic{section}.\arabic{equation}{\rm d}}
\be
X \rfloor y_i \omega = X_i \omega + (-1)^{(p(i) +1)(p(X)+1)} y_i (X \rfloor \omega), \qquad i = 1,\ldots, n,
\ee These relations imply, like in \S~3, that\setcounter{equation}{18}
\renewcommand{\theequation}{\arabic{section}.\arabic{equation}}
\be
X \rfloor d (f) = X(f), \qquad \forall \ f \in {\cal R} = R\langle x
\rangle ,
\ee \be
X \rfloor \omega_1 \omega_2 = (X \rfloor \omega_1) \omega_2 + (-1)^{p(\omega_{1}) (p(X)+1)} \omega_1 (X \rfloor \omega_2), \qquad \forall \ \omega_1, \omega_2 \in \Omega^*,
\ee 
\be X(\omega) = (-1)^{p(X)} d (X \rfloor \omega) + X \rfloor d (\omega), \qquad \forall \ \omega \in \Omega^*.
\ee 
\setcounter{equation}{22}

\noindent
{\bf Example 4.22.} The dif\/ferential $d: \Omega^* \rightarrow \Omega^*$ is an odd derivation, and \be
d \rfloor \omega_1 \omega_2 = (d \rfloor \omega_1) \omega_2 + \omega_1 (d \rfloor \omega_2), \qquad \forall \ \omega_{1,2} \in \Omega^*,
\ee \be
d \rfloor \omega = \mbox{deg}_y (\omega) \omega, \qquad \forall \
\omega  \in \Omega^*.
\ee 
Formula (3.28) has the following ${\bf Z}_2$-graded version:\be
Z_1 \rfloor Z_2 \rfloor \omega = (-1)^{(p(Z_1)+1)(p(Z_2)+1)} (Z_2 \rfloor Z_1 \rfloor \omega), \qquad \forall \ Z_{1,2} \in
\mbox{Der} (\Omega^*, d).
\ee 
Formulae (3.41) and (3.39) now become, respectively:\be
Z \rfloor X(\omega) = (-1)^{p(X)(p(Z)+1)} X(Z \rfloor \omega) + (-1)^{p(X)} [Z, X] \rfloor \omega,
\ee 
\be
\ba{l}\ds (-1)^{p(X)(\ell + \sum\limits^{\ell}_{1} p(Z_{i}))}
 (Z_\ell \rfloor \ldots \rfloor Z_1 \rfloor X(\omega)) = X (Z_\ell \rfloor \ldots \rfloor Z_1  \rfloor \omega) \\[3mm]\ds \qquad + \sum\limits^\ell_{\alpha = 1}
(-1)^{p(X)(\ell - \alpha + \sum\limits_{j \geq \alpha} p(Z_{j}))} (Z_\ell \rfloor \ldots \hat Z_\alpha [Z_\alpha, X] \rfloor
\ldots  \rfloor Z_1 \rfloor \omega).
\ea
\ee 

Finally, formula (3.48) turns into\be
\ba{l}
\ds Z_\ell \rfloor \ldots \rfloor Z_1 \rfloor d (\omega) =
(-1)^{\sum\limits^{\ell}_{ 1}(p (Z_{j})+1)} d (Z_\ell \rfloor \ldots \rfloor Z_1 \rfloor \omega) \\[2mm]\ds \qquad + \sum\limits^\ell_{\alpha =1} (-1)^{u(\alpha)} Z_\alpha (Z_\ell
\rfloor \ldots \hat Z_\alpha \ldots Z_1 \rfloor \omega)  \\[4mm]
\ds \qquad +\sum\limits_{\alpha <
\beta \leq \ell} (-1)^{v (\alpha, \beta)} (Z_\ell \rfloor \ldots \hat Z_\beta [Z_\beta, Z_\alpha] \rfloor \ldots \hat Z_\alpha \ldots Z_1 \rfloor \omega),
\ea
\ee where\setcounter{equation}{28}
\renewcommand{\theequation}{\arabic{section}.\arabic{equation}{\rm a}}
\be
u (\alpha) = \sum_{s < \alpha} (p (Z_s)+1) + p (Z_\alpha) \sum_{j > \alpha} (p (Z_j) + 1),
\ee 
\setcounter{equation}{28}
\renewcommand{\theequation}{\arabic{section}.\arabic{equation}{\rm b}}\be
v (\alpha, \beta) = \sum_{s < \alpha} (p (Z_s)+1) + p (Z_\alpha )\left(1 + \sum_{\alpha < j < \beta} (p( Z_j ) + 1) \right),
\ee with the understanding that empty sums contribute nothing, and that for $\ell = 1$ formula (4.28) becomes simply
\setcounter{equation}{29}
\renewcommand{\theequation}{\arabic{section}.\arabic{equation}}\be
Z_1 \rfloor d (\omega) = Z_1 (\omega) + (-1)^{p (Z_1)+1} d (Z_1 \rfloor \omega),
\ee which is just the formula (4.21).
\medskip

\noindent
{\bf Remark 4.31.}  Nothing is sacred about the ${\bf Z}_2$-grading.  We could easily replace the grading group ${\bf Z}_2$ by an arbitrary
abelian group $\Gamma$.  In the {\it commutative} case, related
calculations can be  found in [9].

\setcounter{section}{5}
\setcounter{equation}{0}
\renewcommand{\theequation}{\arabic{section}.\arabic{equation}}

\subsection*{\S~5. h-Quantum spaces}
Let $R$ continue as the coef\/f\/icient ring.  It is not important what
$R$ really is as long as it is a ${\bf{Q}}$-algebra.  Let $h$ be a
formal parameter commuting with everything.  We shall denote by $R_h$
either $R[h]$ or $R[[h]]$, depending upon the circumstances.  Let
\be R\langle p, q\rangle  = R_h\langle p_1,\ldots, p_n, q_1,\ldots,
q_n\rangle
\ee 
be the ring of polynomials subject to the relations\be
[p_i, p_j] = [q_i, q_j] = 0, \qquad [p_i, q_j] = h \delta_{ij}, \qquad 1 \leq i, j \leq n.
\ee This is our quantum algebra, $-$ or space on which this algebra
serves as the algebra of functions.  Let us consider dif\/ferential
forms on this space.

Let $H \in R\langle p, q \rangle $ be a Hamiltonian.  We have seen in
the preceding paper [11] that even though the $p$'s and
the $q$'s do not commute, there exist the well-def\/ined objects\be
{\partial H \over \partial p_i}, \quad {\partial H \over \partial
q_i}, \qquad  i = 1,\ldots, n,
\ee and that the corresponding partial derivatives commute:\be
{\partial ^2 H \over \partial u_\alpha \partial u_\beta} = {\partial^2
H \over \partial u_\beta \partial u_\alpha}, \qquad u _{\alpha, \beta}
\in \{p_1,\ldots, p_n, q_1,\ldots, q_n \}.
\ee
Thus, we can def\/ine the dif\/ferential $d$ on $R<p,q>$ by setting\be
d (H) = \sum_i \left(d p_i {\partial H \over \partial p_i} + d q_i {\partial H \over \partial q_i} \right).
\ee Alternatively, we can proceed in the spirit of \S~2, and def\/ine the dif\/ferential $d$ to be a derivation of $R\langle p,q\rangle $ with values in\be
\Omega^* = R_h \langle p, q, \overline p, \overline q \rangle ,
\ee with\be
d (p_i) = \overline p_i, \qquad  d (q_i) = \overline q_i, \qquad
i = 1,\ldots, n.
\ee Finally, to set the $d$-complex in $\Omega^*$, we can use the device of \S~3 and set the commutation relations\setcounter{equation}{7}
\renewcommand{\theequation}{\arabic{section}.\arabic{equation}{\rm a}}
\be
d p_i = p_i d + \overline p_i, \qquad d \overline p_i = - \overline p_i d, \qquad i = 1,\ldots, n,
\ee \setcounter{equation}{7}
\renewcommand{\theequation}{\arabic{section}.\arabic{equation}{\rm b}}
\be
d q_i = q_i d + \overline q_i, \qquad d \overline q_i = - \overline q_i d, \qquad i = 1,\ldots, n,
\ee \setcounter{equation}{7}
\renewcommand{\theequation}{\arabic{section}.\arabic{equation}{\rm c}}
\be
d r = rd, \qquad d (r) = 0, \qquad r \in R.
\ee These are previously the commutation relations (2.5).  Since our ring
$R\langle p,q \rangle $ is not free  noncommutative anymore, having
the quantum commutation relations (5.2) imposed upon it, we have to add the corresponding commutation relations on the dif\/ferential $\overline p_i$'s and $\overline q_i$'s.  In view of formulae (5.3)--(5.5), we set \setcounter{equation}{8}
\renewcommand{\theequation}{\arabic{section}.\arabic{equation}}
\be
[\overline p_i, p_j] = [ \overline p_i, q_j] = [ \overline q_i, p_j ]
= [ \overline q_i, q_j] = 0, \qquad 1 \leq i, j \leq n,
\ee \be
[\overline p_i, \overline p_j]_+ = [\overline p_i, \overline q_j]_+
= [ \overline q_i, \overline q_j]_+ = 0, \qquad 1 \leq i, j \leq n,
\ee where\be
[u, v]_+ = uv + vu
\ee is the anti-commutator.  We need only to make sure that the old relations (5.2) in $R\langle p, q \rangle $ and the new ones (5.9), (5.10) in $\Omega^*$ are compatible, but this is obvious once we apply the dif\/ferential $d$ to the relations (5.2).

If we now try to establish a homotopy formula, we quickly discover that this can't be done, since some of the relations (5.2) are not
homogeneous and thus preclude the def\/inition of the dual contraction
$A_t$.  What to do?

Consider the rind of {\it symbols} $R_h [p,q] = R_h [p_1,\ldots,p_n, q_1,\ldots, q_n]$, where the $p$'s and the $q$'s {\it commute}.  Let us agree  to write every polynomial in this ring in the {\it normal form}, with every monomial written as\be
r q^{\ldots}_1 q_2^{\ldots} \ldots p_1^{\ldots} \ldots
p_n^{\ldots}, \qquad r \in R_h.
\ee 

We can also agree  to use the same arrangement of ``{\it normal quantization}'' in the quantum ring $R_h \langle p,q\rangle$.
Upon this agreement, we see  that \begin{enumerate}
\item[(A)] The quantum ring $R_h \langle p, q\rangle $ and the classical ring $R_h [p,q]$ are {\it isomorphic} as f\/iltered {\it vector  spaces} over $R_h$; and,
\item[(B)] With such vecor-space isomorphism at hand, the dif\/ferential $d$ acts in an identical way on both $R_h \langle p, q\rangle$
and $R_h [p, q]$; therefore,
\item[(C)] If we also arrange the ${\bf Z}_2$-graded rings
$\Omega^* R_h \langle p,q\rangle$ and $\Omega^* R_h [p, q]$ into normal forms, the dif\/ferential $d$ will act in an identical way on both of these rings; and thus,
\item[(D)] The de Rham cohomologies of the quantum space are exactly the same as those of the classical one.\end{enumerate}

But the quantum ring $R_h \langle p, q\rangle $ has its uses as the
fundamental building object possessing quantum dif\/ferential forms.
This will be seen in \S~7.
\medskip

\setcounter{equation}{13}

\noindent
{\bf Remark 5.13.} The same rigidity of the cohomologies can be seen in the more general situation outlined in [11] where
the quantum commutation relations (5.2) are replaced by the commutation relations \be
[u_i, u_j] = hc_{ij}, \qquad c_{ij} = - c_{ji} \in {\cal Z} (R)_{h},
\qquad 1 \leq i, j \leq m,
\ee in the ring $R_h \langle u_1,\ldots, u_m\rangle $;
here ${\cal Z} (R)$ is the center of the ring $R$.  The commutation relations (5.9), (5.10) on the dif\/ferentials are replaced by the commutation relations\be
[du_i, u_j] = 0, \qquad 1 \leq i, j \leq m,
\ee \be
[du_i, du_j]_+ = 0, \qquad 1 \leq i, j \leq m.
\ee 

\setcounter{section}{6}
\setcounter{equation}{0}
\renewcommand{\theequation}{\arabic{section}.\arabic{equation}}

\subsection*{\S~6.  Quantum Clebsch representations}

Let ${\cal G}$ be a Lie algebra and $\chi: {\cal G}
 \rightarrow End (V)$ its
representation.  In Classical mechanics, the symplectic space $V
 \oplus V^*$ serves as a symplectic model for the Poisson spaces
 $C^\infty ({\cal G}^*)$ and $C^\infty(({\cal G} \ltimes V)^*)$, where
 ${\cal G} \ltimes V$ is the semidirect sum of ${\cal G}$ and $V$
 w.r.t. the representation
$\chi:\  {\cal G} \ltimes V$ is
the vector  space ${\cal G} \oplus V$ with the commutator
\be
\left[ \left(\matrix{g_1 \cr v_1 \cr} \right), \left(\matrix{g_1 \cr
 v_2 \cr} \right) \right] =  \left(\matrix{[g_1, g_2] \cr\chi (g_1)
 (v_2) - \chi (g_2) (v_1) \cr} \right), \qquad g_{1,2} \in
{\cal G}, \quad  v_{1,2} \in V.
\ee
With suitable modif\/ications, the similar picture persists in
 Classical f\/luid dynamics, with vector spaces being replaced by
 dif\/ferential algebras (see  [10].)
A close look at the Poisson map $C^\infty ((
{\cal G} \ltimes V)^*) \rightarrow C^\infty (V \oplus V^*)$, called
nowadays the {\it Clebsch representation,} shows that it is linear
and quadratic in its arguments, and is thus likely to represent the
Classical remnant of a more general Quantum map.  This is indeed the
case, at least for systems with f\/inite number of degrees of freedom.  Let us see  the details.

Let $\{e_i\}$ be a basis of ${\cal G}$, and $\{f_\alpha\}$ be a basis of $V$.  Let $(A^\beta_{i \alpha})$ be the set of the matrix elements of the representation $\chi$ on $V$:\be
\chi (e_i) (f_\alpha) = \sum_\beta A^\beta_{i \alpha} f_\beta.
\ee The condition on $\chi$ to be a representation,
\be
\chi ([g_1, g_2]) = [\chi (g_1), \chi(g_2)], \qquad \forall \ g_1, g_2
\in {\cal G}, \ee 
translates into the set of equalities\be
\sum_k c^k_{ij} A^\gamma _{k \alpha} = \sum_\beta \left(A^\gamma_{i \beta} A^\beta_{j \alpha} - A^\gamma _{j \beta} A^\beta_{i \alpha}\right),
\ee where $\{c_{ij}^k\}$ are the structure constants of ${\cal G}$ in the basis $\{e_i\}$: \be
[e_i, e_j] = \sum_k c^k_{ij} e_k.
\ee All our constants are from $R$ which is now assumed to be {\it
commutative}.
Let $\{g^\beta\}$ be the dual basis in $V^*$.  Let $\nabla: V \otimes
V^* \rightarrow  {\cal G}^*$ be the basic Clebsch map of Chapter 8 in
[10], def\/ined by the formula\be
\langle  \nabla (v \otimes v^*), g \rangle  =
\langle  v^*, \chi (g) (v) \rangle ,
\ee so that, in components,\setcounter{equation}{6}
\renewcommand{\theequation}{\arabic{section}.\arabic{equation}{\rm a}}
\be
f_\alpha \nabla g^\beta = \nabla (f_\alpha \otimes g^\beta) = \sum_i A^\beta_{i \alpha} e^i \qquad \Leftrightarrow
\ee \setcounter{equation}{6}
\renewcommand{\theequation}{\arabic{section}.\arabic{equation}{\rm b}}
\be
(f_\alpha \nabla g^\beta)_i = A^\beta_{i \alpha}.
\ee 
\setcounter{equation}{8}
\renewcommand{\theequation}{\arabic{section}.\arabic{equation}}

\noindent
{\bf Lemma 6.8 (Quantum Clebsch representation.)} {\it Let $\{F^\alpha; G_\alpha\}$ be the generators of the Quantum algebra
$R_h \langle F, G\rangle $, with the commutation relations\be
[F^\alpha, F^\beta] = [G_\alpha, G_\beta] = 0, \qquad [F^\alpha, G_\beta] = h \delta^\alpha_{\beta}.
\ee Set\be
e_i = \sum_{\alpha \beta} A^\beta_{i \alpha} F^\alpha G_\beta h^{-1}, \ee 
\be
f_\alpha = k G_\alpha h^{-1}, \qquad k \in R.
\ee Then the thus def\/ined elements satisfy the commutation relations of the basis in ${\cal G}$ and in ${\cal G} \ltimes V:$\be
[e_i, e_j] = \sum_k c^k_{ij} e_k,
\ee \be
[e_i, f_\alpha] = \sum_\beta A^\beta_{i \alpha } f_\beta.
\ee 
\be
[f_\alpha, f_\beta] = 0.
\ee 
}
\noindent
{\bf Proof.}  We have,\be
[e_i, e_j] \ {\mathop{=}\limits^{\mbox{\scriptsize \rm [by (6.10)]}}}
\  \sum h^{-2} A^\beta_{i \alpha} A^\nu_{j \alpha}
[F^\alpha G_\beta, F^\mu G_\nu].
\ee Now,\be
[F^\alpha G_\beta, F^\mu G_\nu] =
h\left(- \delta^\mu_{\beta} F^\alpha G_\nu+ \delta^\alpha_{\nu} F^\mu G_\beta\right).
\ee Indeed,\[
\ba{l}
\ds
[F^\alpha G_\beta, F^\mu G_\nu] = F^\alpha [G_\beta, F^\mu G_\nu] + [F^\alpha, F^\mu G_\nu] G_\beta \\[2mm]\ds \qquad =F^\alpha [G_\beta, F^\mu] G_\nu + F^\mu [F^\alpha, G_\nu]
G_\beta \ {\mathop{=}\limits^{\mbox{\scriptsize \rm [by (6.9)]}}} \
 \left(- \delta^\mu_\beta F^\alpha G_\nu + \delta^\alpha_\nu F^\mu G_\beta\right) h. \ea
\]
Substituting (6.16) into (6.15), we f\/ind:\[
\ba{l}
\ds [e_i, e_j] = \sum h^{-2} A^\beta_{i \alpha} A^\nu_{j \alpha} h
\left(-\delta^\mu _\beta F^\alpha G_\nu + \delta^\alpha_\nu F^\nu
G_\beta\right)
\\[4mm]
\ds \qquad = - h^{-1} \sum A^\nu_{j \gamma} A^\gamma_{i \alpha}
F^\alpha G_\nu + h^{-1} \sum A^\nu_{i \gamma} A^\gamma_{j \alpha}  F^\alpha G_\nu \\[4mm]
\ds \qquad = h^{-1} \sum_{\alpha \nu} F^\alpha G_\nu \sum_\gamma
\left(A^\nu_{i \gamma} A^\gamma_{j \alpha} - A^\nu_{j \gamma}
A^\gamma_{i\alpha}\right) \\[4mm]
\ds \qquad {\mathop{=}\limits^{\mbox{\scriptsize \rm [by (6.4)]}}} \
 h^{-1} \sum_{\alpha \nu} F^\alpha G_\nu \sum_k c^k_{ij} A^\nu_{k \alpha} = \sum_k c^k_{ij} \sum_{\alpha \nu} h^{-1} A^\nu_{k \alpha } F^\alpha G_\nu \
{\mathop{=}\limits^{\mbox{\scriptsize \rm [by (6.10)]}}} \ \sum_k c^k_{ij} e_k,
\ea
\]and this is formula (6.12).

Next, \[
\ba{l}
\ds [e_i, f_\alpha] \
{\mathop{=}\limits^{\mbox{\scriptsize \rm [by (6.10), (6.11)]}}} \
\sum h^{-2} A^\nu_{i \mu} k [F^\mu G_\nu, G_\alpha] \\[3mm]
\ds \qquad = \sum h^{-2} A^\nu_{i \mu} k \delta^\nu_\alpha h G_\nu = \sum A^\nu_{i \alpha} k G_\nu h^{-1} \
{\mathop{=}\limits^{\mbox{\scriptsize \rm [by (6.11)]}}} \ \sum A^\nu_{i \alpha} f_\nu,
\ea
\]and this is formula (6.13).

Formula (6.14) is obvious. \qquad \rule{2mm}{4mm}

\setcounter{equation}{17}

\medskip

\noindent
{\bf Remark 6.17.} The Quantum Clebsch formulae (6.10), (6.11) are {\it singular} in $h$ and thus do not allow the passage to the
quasiclassical limit.  To make sure such passage is possible, we should
rescale these formulae into the form:\be
\bar e_i = \sum A^\beta_{i \alpha} F^\alpha G_\beta,
\ee \be
\bar f_\alpha = k G_\alpha,
\ee 
\be
[\bar e_i, \bar e_j] = h \sum c^k_{ij} \bar e_k,
\ee \be
[\bar e_{i}, \bar f_\alpha] = h \sum A^\beta_{i \alpha} \bar f_\beta, \ee 
\be
[\bar f_\alpha, \bar f_\beta] = 0.
\ee In the limit $h \rightarrow 0$, $\bar e_i$'s and $\bar f_\alpha$'s form the polynomial generators of the Poisson function rings $R[{\cal G}^*]$
and $R[({\cal G }\ltimes V)^*]$.
\medskip

\noindent
{\bf Remark 6.23.} In the older literature, the Quantum Clebsch
representations, considered primarily for real and complex semisimple
Lie algebras, have been called ``canonical realizations'' of Lie
algebras (see , e.g., [13, 7, 4, 2]); a
more recent terminology is ''boson representations''.  Since there
exist also the so-called ``boson-fermion representations'', one can
suspect that Quantum Clebsch representations can be generalized to
include fermions.  This is
indeed the case.  Formulae (6.10)--(6.15) and (6.16) remain unchanged
if
formulae (6.9) are replaced by the formulae \setcounter{equation}{23}
\renewcommand{\theequation}{\arabic{section}.\arabic{equation}{\rm a}}
\be
F^\alpha F^\beta - (-1)^{p(\alpha) p(\beta)} F^\beta F^\alpha = 0, \ee 
\setcounter{equation}{23}
\renewcommand{\theequation}{\arabic{section}.\arabic{equation}{\rm b}}
\be
G_\alpha G_\beta - (-1)^{p(\alpha) p(\beta)} G_\beta G_\alpha = 0, \ee 
\setcounter{equation}{23}
\renewcommand{\theequation}{\arabic{section}.\arabic{equation}{\rm c}}
\be
F^\alpha G_\beta - (-1)^{p(\alpha) p(\beta))} G_\beta F^\alpha = (-1)^{p (\alpha)} \delta^\alpha_\beta h,
\ee and formula (6.14) is replaced by formula\setcounter{equation}{24}
\renewcommand{\theequation}{\arabic{section}.\arabic{equation}}
\be
f_\alpha f_\beta - (-1)^{p (\alpha) p (\beta)} f_\beta f_\alpha = 0; \ee 
here\be
p(\alpha) = p(F^\alpha) = p(G_\alpha)
\ee are aribtrary ${\bf{Z}}_2$-gradings on the space of Quantum variables
$\{F^\alpha\}$ and $\{G_\beta\}$, distinguishing bosons (with $p(\alpha) = 0)$ from fermions (with $p (\alpha) = 1$).  The details are left to the reader.

The Quantum Clebsch representations will be used in the next Section
to construct a complex of dif\/ferential forms on the Universal
enveloping algebra $U({\cal G}$).
\setcounter{equation}{27}

\medskip

\noindent
{\bf Remark 6.27.} The Quantum Cebsch representation constructed in this Section is {\it general}, i.e., not dependent upon any particular properties of the Lie algebra ${\cal G}$.  When one considers some
{\it special} Lie algebras, one can naturally expect some extra
ef\/fects.  For example, for the quantum group $GL(V)$, acting on a pair of vector spaces $V$ and $V^*$ by the rule\be
\pmx^\prime = M \pmx, \qquad x^\prime_i = \sum_\alpha M_{i \alpha}
x_\alpha,
\ee 
\be\pmp^{t \prime} = \pmp^t M, \qquad p^\prime_i = \sum_\alpha p_\alpha
M_{\alpha i}, \ee 
with the commutation relation on $V$ and $V^*$ given by the generalized commutation relations of the form\be
\sum_{k \ell} R^{k \ell}_\alpha x_k x_\ell = 0, \qquad \alpha \in
{\cal A},\ee 
\be
\sum_\beta \overline R_\beta^{k \ell} p_k p_\ell = 0, \qquad \beta \in {\cal B},
\ee where ${\cal A}$ and ${\cal B}$ are some index sets, the induced
quantum group structure on $GL(V)$ is easily seen to allow the representation\be
M_{i \alpha} = u_i v_\alpha,
\ee where the $u$'s and the $v$'s satisfy the commutation relations\be
\sum_{k \ell} R^{k \ell}_\alpha u_k u_\ell = 0, \qquad \alpha \in
{\cal A},\ee 
\be
\sum_{k \ell} \overline R^{k \ell}_\beta v_k v_\ell = 0, \qquad \beta \in {\cal B},
\ee \be
[u_k, v_\ell] = 0, \qquad \forall \ k, \ell,
\ee We shan't pursue this avenue further.

\setcounter{section}{7}
\setcounter{equation}{0}
\renewcommand{\theequation}{\arabic{section}.\arabic{equation}}

\subsection*{\S~7.  Dif\/ferential forms on Lie algebras}
Continuing with the notation of the preceding Section, let $U ({\cal
G})$ be the universal enveloping algebra of the Lie algebra ${\cal
G}$.  This is simply the noncommutative ring $R\langle  e_1,\ldots, \rangle$, subject to the relations\be
e_i e_j - e_j e_i = \sum_k c^k_{ij} e_l, \qquad \forall \ i, j.
\ee 
We wish to construct an analog of the ring of dif\/ferential forms $\Omega^*$ for $U({\cal G})$, preferably on the lines of \S~2.  In order to
achieve, this, we need to determine the commutation relations between
the $e_i$'s, and $de_j$'s, of the form \be
[e_i, de_j] = \sum_k \theta_{ij}^k de_k, \qquad \forall \ i, j. \ee 
To be consistent with the Lie algebra structures (7.1), the relations (7.2) have to be compatible with the relations\be
[de_i, e_j] + [e_i, de_j] = \sum_k c^k_{ij} de_i.
\ee This amounts to the series of the identities\be
\theta_{ij}^k - \theta^k_{ji} = c^k_{ij}, \qquad \forall \ i,j, k.
\ee 
Also, formuale (7.2) must def\/ine a {\it representation} of the Lie algebra ${\cal G}$ on the vector space of dif\/ferentials $\{d e_i\}$; by formula (6.4), this amounts to the series of identities\be
\sum_{s} \left(\theta^k_{is} \theta^s_{j\ell} -
\theta^k_{js} \theta^s_{i\ell}\right) = \sum_s c^s_{ij} \theta^k_{s \ell} , \qquad \forall \ i,j,k,\ell.
\ee 
These are to be compared with the Jacobi identity for the structure constants $c^k_{ij}$'s:\be
\sum_i \left(c^k_{is} c^s_{j \ell} - c^k_{js} c^s_{i\ell}\right)
= \sum_s c^s_{ij}  c^k_{s\ell}, \qquad \forall \ i,j,k,\ell.
\ee 
Clearly, such structure consists $\theta^k_{ij}$'s do not exist in {\it general}, although they may and do in fact exist in {\it particular} (see  \S\S~8,~9).
Let us bring in the Quantum Clebsch representation of the preceding Section.  Thus, we abandon our initial goal to have a dif\/ferential-forms-complex solely in terms of the Lie algebra ${\cal G}$ and use the additional
data in the form of a representation $\chi$ of ${\cal G}$ on a vector
space $V$.
By formula (6.10),\be
e_i = h^{-1} \sum_{\alpha \beta} A^\beta_{i \alpha} F^\alpha G_\beta.
\ee Hence, we can set\be
de_i = h^{-1} \sum A^\beta_{i \alpha} \left(d F^\alpha G_\beta + F^\alpha d G_\beta\right).
\ee Denoting\be
\omega_\beta^\alpha = h^{-1}dF^\alpha G_\beta, \qquad \Omega^\alpha_\beta= h^{-1} F^\alpha d G_\beta,
\ee we get\be
d e_i = \sum A^\beta_{i \alpha} \left(
\omega^\alpha_\beta + \Omega^\alpha_\beta\right). \ee 
Now,\[
\ba{l}
\ds \left[e_i, \omega^\beta_\alpha\right] = \left[h^{-1} \sum_{\mu \nu}
A^\nu_{i \mu} F^\mu G_\nu, \ h^{-1} d F^\alpha G_\beta\right] \ {\mathop{=}\limits^{\mbox
{\scriptsize \rm [by (5.9)]}}} \ h^{-2} \sum A^\nu_{i \mu} d F^\alpha [F^\mu, G_\beta] G_\nu \\[4mm]\ds \qquad =
h^{-1} \sum_\nu A^\nu_{i \beta} d F^\alpha G_\nu = \sum_\nu A^\nu_{i \beta}\omega^\alpha_{\nu } \qquad \Rightarrow
\ea
\]
\be[e_i, \omega^\alpha_\beta] = \sum_\nu A^\nu_{i \beta} \omega^\alpha_\nu. \ee 
Similarly,\[
\ba{l}
\ds [e_i, \Omega^\alpha_\beta] = \sum h^{-2} A^\nu_{i \mu} [F^\mu G_\nu, F^\alpha d G_\beta] = \sum h^{-2} A^\nu_{i \mu} F^\nu [G_\nu, F^\alpha] d G_\beta \\[4mm]\ds \qquad = - h^{-1} \sum A^\alpha_{i \mu} F^\mu dG_\beta = -
\sum_\mu A^\alpha _{i \mu} \Omega^\mu_\beta \qquad \Rightarrow
\ea
\]\be
[e_i, \Omega^\alpha_\beta] = - \sum_\mu A^\alpha_{i \mu}
\Omega^\mu_{\beta}.
\ee Combining formulae (7.10)--(7.12), we f\/ind that\be
[e_i, d e_j] = \sum_{\alpha \beta} \left< \left(\sum_\gamma A^\beta_{i \gamma} A^\gamma_{j \alpha} \right) \omega^\alpha_{\beta} - \left(
 \sum_\gamma A^\beta_{j \gamma} A^\gamma_{i \alpha} \right)
\Omega^\alpha_\beta \right>, \ee 
still another indication that our original goal of constructing the
dif\/ferential complex on $\Omega^* U ({\cal G})$ was ill-posed.  (If
${\cal G}$ issemisimple or reductive, we can chose some special representation:  adjoint, coadjoint, fundamental, etc.  But these fall under ``special'' category.  On the other hand, we are interested in a general construction.)

Let us verify that the dif\/ferential relations (7.10)--(7.12) are compatible with the Lie algebra relations (7.1).  We have to check the identity\be
[d e_i, e_j] + [e_i, d e_j] = \sum_k c^k_{ij} d e_k.
\ee By formula (7.13), for the LHS of formula (7.14) we get\[
\hspace*{-10pt}
\ba{l}
\ds [de_i, e_j] + [e_i, de_j] = \sum_{\alpha \beta} \left< \sum_{\gamma }\left(A_{j \gamma}^\beta A^\gamma_{j \alpha} - A^\beta_{j \gamma}
A^\gamma_{i \alpha}\right) \omega^\alpha_{\beta} -
\sum_\gamma \left(A^\beta_{j \gamma} A^\gamma_{i \alpha} - A^\beta_{i \gamma} A^\gamma_{j \alpha}\right)
 \Omega^\alpha_\beta \right>
\\[4mm]\ds = \sum_{\alpha \beta} \left< \sum_\gamma \left(A^\beta_{i \gamma} A^\gamma_{j \alpha} - A^\beta_{j \gamma} A^\gamma_{i \alpha} \right)
\left( \omega^\alpha_{\beta} + \Omega^\alpha _\beta \right)
\right> \ {\mathop{=}\limits^{\mbox{\scriptsize \rm [by (6.4)]}}}
\  \sum_{\alpha \beta} c^k_{i j} A^\beta_{k \alpha} \left(\omega^\alpha_\beta + \Omega^\alpha_{\beta} \right) \\[5mm]
\ds \qquad {\mathop{=}\limits^{\mbox{\scriptsize \rm [by (7.4)]}}} \
 \sum_k c^k_{ij} d e_k,
\ea
\]and this is the RHS of formula (7.14).

The construction of our dif\/ferential complex is not complete yet, for we have to def\/ine the action of the dif\/ferential $d$ on the generators
$\omega ^\alpha_\beta$ and $\Omega^\alpha_\beta$.  Keeping our
deep-background formulae (7.9) in mind, we see  that in the Quantum
Clebsch representation we should take\be
d (\omega^\alpha_\beta) = - h^{-1} dF^\alpha d C_\beta, \qquad d (\Omega^\alpha_\beta) = h^{-1} dF^\alpha d G_\beta.
\ee Accordingly, we introduce {\it new generators} $\rho^\alpha_\beta$ into $\Omega^*$, and set\be
d (\omega^\alpha_\beta) = - \rho^\alpha_{\beta},
\ee \be
d (\Omega^\alpha_\beta) = \rho^\alpha_\beta,
\ee \be
d (\rho^\alpha_\beta) = 0.
\ee To keep track of the dif\/ferential degrees, let us set \setcounter{equation}{18}
\renewcommand{\theequation}{\arabic{section}.\arabic{equation}{\rm a}}
\be
p_y (e_i) = p_y (R) = 0,
\ee \setcounter{equation}{18}
\renewcommand{\theequation}{\arabic{section}.\arabic{equation}{\rm b}}
\be
p_y (\omega^\alpha_\beta) = p_y (\Omega^\alpha_\beta) = 1,
\ee \setcounter{equation}{18}
\renewcommand{\theequation}{\arabic{section}.\arabic{equation}{\rm c}}
\be
p_y (\rho^\alpha_\beta) = 2.
\ee These gradings make the dif\/ferential $d$ into a homogeneous operator of $p_y$-degree  1; the ${\bf Z}_2$-grading on $\Omega^*$ is given, as usual, by the elements $p_y$ ($\mbox{mod}\; 2$).  Formulae (7.10) and
(7.16)--(7.18) show that $d^2=0$ on $\Omega^*$.  However, we still
have to verity that our operator $d$ preserves the commutation relations (7.11), (7.12), and some relations still to come, such as
\setcounter{equation}{19}
\renewcommand{\theequation}{\arabic{section}.\arabic{equation}}\be
[e_i, \rho^\alpha_{\beta}] = 0.
\ee 
Applying the dif\/ferential $d$ to the relation (7.11), rewritten as\[
e_i \omega^\alpha_\beta - \omega^\alpha_\beta e_i = \sum_\nu A^\nu_{i \beta} \omega^\alpha_\nu,
\]we f\/ind\be
\sum_{\mu \nu} A^\nu_{i \mu} (\omega^\mu_\nu + \Omega^\mu_\nu) \omega^\alpha_\beta - e_i \rho^\alpha_\beta + \rho^\alpha_\beta e_i + \omega^\alpha_\beta \sum_{\mu \nu} A^\nu_{i \mu} (\omega^\mu_{\nu} + \Omega^\mu_\nu ) = - \sum_\nu A^\nu_{i \beta} \rho^\alpha_\nu.
\ee Remembering the nature of $\rho^\alpha_\beta$  as $h^{-1} d F^\alpha G_\beta$, let us postulate that the $\rho^\alpha_\beta$'s commute with everything:\setcounter{equation}{21}
\renewcommand{\theequation}{\arabic{section}.\arabic{equation}{\rm a}}
\be
[\rho^\alpha_{\beta}, e_i] = 0,
\ee 
\setcounter{equation}{21}
\renewcommand{\theequation}{\arabic{section}.\arabic{equation}{\rm b}}\be
[\rho^\alpha_\beta, \omega^\mu_\nu] = 0,
\ee 
\setcounter{equation}{21}
\renewcommand{\theequation}{\arabic{section}.\arabic{equation}{\rm c}}\be
[\rho^\alpha_\beta, \Omega^\mu_\nu] = 0,
\ee 
\setcounter{equation}{21}
\renewcommand{\theequation}{\arabic{section}.\arabic{equation}{\rm d}}\be
[\rho^\alpha_\beta, \rho^\mu_\nu] = 0.
\ee Obviously, these relations remain consistent when acted upon by the
dif\/ferential $d$.  And while we are at it, we can make use of the
def\/ining background relations (7.9) and postulate the commutation
relations
\setcounter{equation}{22}
\renewcommand{\theequation}{\arabic{section}.\arabic{equation}{\rm a}}
\be
[\omega^\alpha_\beta, \omega^\mu_\nu]_+ = 0,
\ee 
\setcounter{equation}{22}
\renewcommand{\theequation}{\arabic{section}.\arabic{equation}{\rm b}}\be
[\Omega^\alpha_\beta, \Omega^\mu_\nu]_+ = 0,
\ee \setcounter{equation}{22}
\renewcommand{\theequation}{\arabic{section}.\arabic{equation}{\rm c}}
\be
[\omega^\alpha_\beta, \Omega^\mu_\nu]_+ = - \delta^\mu_\beta \rho^\alpha_\nu.
\ee The latter formula is suggested by the following background calculation:\setcounter{equation}{23}
\renewcommand{\theequation}{\arabic{section}.\arabic{equation}}
\be
\ba{l}
[\omega^\alpha_\beta, \Omega^\mu_\nu]_+ = h^{-2} (dF^\alpha G_\beta F^\mu d G_\nu + F^\mu d G_\nu d F^\alpha G_\beta)\\[2mm]\ds \qquad = h^{-2} d F^\alpha d G_\nu (G_\beta F^\mu - F^\mu G_\beta) = - h^{-1} d F^\alpha d G_\nu \delta^\mu_\beta = - \delta^\mu_\beta \rho^\alpha_\nu.
\ea
\ee 
Now, substituting formulae (7.22a), (7.23a) into the identity to be
verif\/ied, (7.21), we get\[
\sum_{\mu \nu} A^\nu_{i \mu} \left(\Omega^\mu_\nu \omega^\alpha_\beta + \omega^\alpha_\beta \Omega^\mu_\nu\right) = - \sum_\nu A^\nu_{i
\beta} \rho^\alpha _\nu,
\]which is true in view of formula (7.23c).  Similarly, applying the dif\/ferential $d$ to formula (7.12), we get\[
[e_{i}, \rho^\alpha_\beta] +
\left[\sum_{\mu \nu} A^\nu_{i \mu} \left(\omega^\mu_\nu + \Omega^\mu_\nu\right), \Omega^\alpha_\beta\right]_+ =
- \sum_\mu A^\alpha_{i \mu} \rho^\mu_\beta,
\]which is true in view of formulae (7.22a), (7.23b), (7.23c).

Finally, applying the dif\/ferential $d$ to the remaining relations (7.23) and using the formula\be
d ([\varphi, \psi]_+) = [d (\varphi), \psi] - [\varphi, d (\psi)],
\qquad p
(\varphi) = p (\psi) = 1 \in {\bf Z}_2,
\ee we see  that the resulting relations are satisf\/ied in view of formulae (7.22), (7.16)--(7.18).
The end result is the $d$-complex $\Omega^*$, with the generators $\{e_i\}$, $\{\omega^\alpha_\beta\}$, $\{\Omega^\alpha_\beta\}$,
$\{\rho^\alpha _\beta\}$, the relations (7.1), (7.11), (7.12),
(7.22), (7.23), and the action of the dif\/ferential $d$ given by the formulae (7.10), (7.16)--(7.18).  The Quantum generators $\{F^\alpha\}$ and $\{G_\alpha\}$, having served their suggestive purpose, do not enter into the picture anymore.  But they still can be of some use:  note that our complex $\Omega^*$ is not f\/inite-dimensional over $U({\cal G})$.  To make it so, we can use formulae (7.9), (7.15) and
impose the additional relations\setcounter{equation}{25}
\renewcommand{\theequation}{\arabic{section}.\arabic{equation}{\rm a}}
\be
\omega^\alpha_\mu \omega^\alpha_\nu = 0,
\ee 
\setcounter{equation}{25}
\renewcommand{\theequation}{\arabic{section}.\arabic{equation}{\rm b}}\be
\Omega^\mu_\beta \Omega^\nu_\beta = 0,
\ee 
\setcounter{equation}{25}
\renewcommand{\theequation}{\arabic{section}.\arabic{equation}{\rm c}}\be
\omega^\alpha_\mu \rho^\alpha_\nu = 0,
\ee \setcounter{equation}{25}
\renewcommand{\theequation}{\arabic{section}.\arabic{equation}{\rm d}}
\be
\Omega^\mu_\beta \rho^\nu_\beta = 0,
\ee \setcounter{equation}{25}
\renewcommand{\theequation}{\arabic{section}.\arabic{equation}{\rm e}}
\be
\rho^\alpha_\mu \rho^\alpha_\nu = 0, \qquad \rho^\mu_\beta \rho^\nu_\beta = 0.
\ee 
These relations are obviously preserved under the action of the
dif\/ferential $d$.  The resulting complex $\{\Omega^{*fin}, d\}$, and
the bigger complex $\{\Omega^*, d\}$, are easily seen to be natural
in the category of ${\cal G}$-modules.  Both these complexes,
suggested by Quantum mechanical considerations, are quite dif\/ferent
from the usual Lie-algebraic ones (see  [5, 6]),
and the low-dimensional cohomologies of the new complexes should have a dif\/ferent interpretation as well.
\medskip

\noindent
{\bf Remark 7.27.} The Quantum Clebsch map (7.7) which is serving as a motivator of the $\{\Omega^*, d\}$-complex, is constructed from elements of both $V$ and $V^*$.  Accordingly, nothing is gained if we replace the representation $\chi$ on $V$ by the dual representation
$\chi^d$ on $V^*$.

In the next three  Sections we shall look at the special Lie algebras $af\!f(1)$, $gl(V)$, and $so(V)$, where the size of the dif\/ferential complex $\{\Omega^*, d\}$ constructed in this Section can be substantially reduced.

\setcounter{section}{8}
\setcounter{equation}{0}
\renewcommand{\theequation}{\arabic{section}.\arabic{equation}}

\subsection*{\S~8.  The Lie algebra {\itshape aff}(1) and its
generalizations}

Let $G = Af\!f(1)$ be the Lie group of af\/f\/ine transformations of
the line,
\be
\{x \mapsto x^\prime = a x + b, \qquad a \ \mbox{is  invertible}.\} \ee 
From the matrix representation
\be
\left(\matrix{x^\prime \cr1 \cr} \right) = \left(\matrix{a & b \cr0 & 1 \cr} \right) \left(\matrix{x \cr1 \cr} \right)
\ee of this Lie group, we can represent the Lie algebra ${\cal G} =
af\!f(1)$ as the subspace in $gl(2)$ of the form\be
af\!f (1) = \left\{ \left(\matrix{* & * \cr0 & 0 \cr} \right) \right\} .
\ee Setting\be
e_1 = \left(\matrix{1 & 0 \cr0 & 0 \cr} \right), \qquad  e_2 = \left(\matrix{0 & 1 \cr 0 & 0 \cr} \right),
\ee we get the commutator in ${\cal G}$\be
[e_1, e_1] = e_2.
\ee The same commutator relation (8.5) is af\/forded by the following generators in the Quantum algebra $R_h \langle p, q\rangle $:\be
E_1 = p, \qquad  E_2 = e^{q/h}.
\ee Since\be
d (E_1) = dp, \qquad d (E_2) = h^{-1} E_2 dq,
\ee we f\/ind\setcounter{equation}{7}
\renewcommand{\theequation}{\arabic{section}.\arabic{equation}{\rm a}}
\be
[d(E_1), E_1] = [d(E_1), E_2] = 0,
\ee 
\setcounter{equation}{7}
\renewcommand{\theequation}{\arabic{section}.\arabic{equation}{\rm b}}\be
[E_1, d (E_2)] = d (E_2), \qquad [E_2, d (E_2)] = 0.
\ee Thus, we can take the relations (8.8) as def\/ing the commutation relations in $\Omega^* (U({\cal G})):$\setcounter{equation}{8}
\renewcommand{\theequation}{\arabic{section}.\arabic{equation}{\rm a}}
\be
[e_1, d e_1] = [e_2, d e_1] = 0,
\ee \setcounter{equation}{8}
\renewcommand{\theequation}{\arabic{section}.\arabic{equation}{\rm b}}
\be
[e_1, d e_2] = d e_2, \qquad [e_2, de_2] = 0 .
\ee To make the combined relations (8.5), (8.9) in $\Omega^*$ self-consistent, we have to apply the dif\/fe\-ren\-tial $d$ to the commutation relations (8.9). We thus obtain:\setcounter{equation}{9}
\renewcommand{\theequation}{\arabic{section}.\arabic{equation}{\rm a}}
\be
(d e_1)^2 = (d e_2)^2 = 0,
\ee 
\setcounter{equation}{9}
\renewcommand{\theequation}{\arabic{section}.\arabic{equation}{\rm b}}\be
(de_1) (de_2) + (de_2) (de_1) = 0.
\ee 
To show that the cohomologies of the constructed complex $\{\Omega^*,
d\}$ are trivial, we could in principle embed the complex
$\Omega^*(U({\cal G}))$ into $\Omega^*(R_h \langle p, q\rangle )$; the latter has been proven
to be trivial in \S~3, but only in the {\it{polynomial}} setting, and
formula (8.6) contains the exponential function; the latter, in
addition, is {\it singular} in $h$.  This is not fatal for the
argument, but it's more ef\/f\/icient to use the {\it method} of \S~3
instead of the f\/inal result.  Namely, let's identify $U({\cal G})$, {\it as a vector space}, with
the polynomial ring $R[e_1, e_2]$ via the {\it normal} ordering of
monomials in the form\setcounter{equation}{10}
\renewcommand{\theequation}{\arabic{section}.\arabic{equation}}
\be
\{r e_1^{\ldots} e_2^{\ldots} \}
\ee Then, by formulae (8.9)\be
d\left(r e^n_1 e^m_2\right) =
 r \left(n e^{n-1}_1 e^m_2 de_1 + me^n_1 e^{m-1}_2 d e_2\right), \ee 
so that $\{\Omega^* (U ({\cal G})), d\}$ is isomorphic, as a vector
space, to $\{\Omega^* (R[e_1, e_2]), d\}$.  Thus, the cohomologies of
both are identical, and trivial.

The example of the Lie algebra $af\!f(1)$ suggests that one may have a
similar result for more general solvable Lie algebras.  Another
generalization, less
sweeping, is to replace our Lie algebra $af\!f(1)$ by the Lie algebra ${\cal G} = {\cal G}(A)$, where $A$ is an arbitrary constant $n
\times n$ matix, and ${\cal G} (A)$ has the generators\be
e_1,\ldots, e_n, \overline e_1,\ldots, \overline e_n,
\ee with the relations
\setcounter{equation}{13}
\renewcommand{\theequation}{\arabic{section}.\arabic{equation}{\rm a}}\be
[e_i, e_j] = [\overline e_i, \overline e_j] = 0, \qquad 1 \leq i, j \leq n, \ee 
\setcounter{equation}{13}
\renewcommand{\theequation}{\arabic{section}.\arabic{equation}{\rm b}}
\be
[e_i, \overline e_j] = A_{ji} \overline e_j, \qquad  1 \leq i, j \leq n. \ee 
This Lie algebra has the Quantum model $R_h \langle p_1,\ldots,
p_n, q_1,\ldots, q_n\rangle$, with\setcounter{equation}{14}
\renewcommand{\theequation}{\arabic{section}.\arabic{equation}}
\be
E_i = p_i, \qquad \overline E_i = \exp \left(h^{-1} \sum_s A_{is} q_s
\right), \qquad 1 \leq i \leq n.
\ee Accordingly, we impose the following commutation relations in $\Omega^*(U({\cal G})):$
\setcounter{equation}{15}
\renewcommand{\theequation}{\arabic{section}.\arabic{equation}{\rm a}}\be
[d e_i, e_j] = [d e_i, \overline e_j] = 0 = [d \overline e_i, \overline e_j], \qquad   1 \leq i, j \leq n,
\ee \setcounter{equation}{15}
\renewcommand{\theequation}{\arabic{section}.\arabic{equation}{\rm b}}
\be
[e_i, d \overline e_j] = A_{ji} d \overline e_j, \qquad 1 \leq i, j
\leq n.
\ee To insure self-consistency betwee n relations (8.14), (8.16), we have
to adjoin the skewsymmetry relations\setcounter{equation}{16}
\renewcommand{\theequation}{\arabic{section}.\arabic{equation}}
\be
[d e_i, d e_j]_+ = [d e_i, d \overline e_j]_+ = [d \overline e_i, d \overline e_j]_+ = 0, \qquad 1 \leq i, j \leq n.
\ee To see  that cohomologies of the constructed complex are trivial, we
identify $U({\cal G})$ with $R[ e_1,\ldots, e_n, \overline
e_1,\ldots, \overline e_n]$ (again, as vector spaces) via the normal
ordering
\be
\{r e_1^{\ldots} \ldots e_n^{\ldots}  \overline e_1^{\ldots}  \ldots \overline e_n^{\ldots}\},
\ee and agree  to write the dif\/ferentials $d(\ldots)$ on the {\it{right}} from the normalized monomials (8.18).  The complex $\{\Omega^* (U
({\cal G})), d\}$ then looks exactly as the one for the commutative
polynomial algebra $R[e_1,\ldots, e_n, \overline e_1,\ldots,
\overline e_n]$.

It remains to give the algebra ${\cal G}(A)$  a suitable monicker.  The traditional method to chose such is to attach to the nameless subject of attention the adjective Schr\"odinger, Heisenberg,
Dirac, etc., but these worthies have already everything under the sun
named after them.  Accordingly, I shall call the algebra ${\cal G}(A)$ the
{\it Ehrenfest algebra}.

\setcounter{section}{9}
\setcounter{equation}{0}
\renewcommand{\theequation}{\arabic{section}.\arabic{equation}}

\subsection*{\S~9.  The Lie algebra {\itshape gl}({\itshape V})}

The Lie algebra $gl(V)$ has two most natural representations:  the natural actions on $V$ and $V^*$.  If we chose a basis $\{f_\alpha \}$ in $V$ and the corresponding basis of elementary matrices $\{e_{ij}\}$ in
$End(V)$, then\be
e_{ij} (f_\alpha) = \delta_{j \alpha} f_i.
\ee Thus, the structure constants entering formulae (6.2) and (7.10) are\be
A^\beta_{ij|\alpha}  = \delta_{j \alpha} \delta^\beta_i.
\ee Accordingly, formulae (7.10)--(7.12) become\setcounter{equation}{2}
\renewcommand{\theequation}{\arabic{section}.\arabic{equation}{\rm a}}
\be
d (e_{ij}) = \omega^j_i + \Omega^j_i,
\ee 
\setcounter{equation}{2}
\renewcommand{\theequation}{\arabic{section}.\arabic{equation}{\rm b}}\be
[e_{ij}, \omega^\alpha_\beta] = \delta_{j \beta} \omega^\alpha_i, \ee 
\setcounter{equation}{2}
\renewcommand{\theequation}{\arabic{section}.\arabic{equation}{\rm c}}
\be
[e_{ij}, \Omega^\alpha_\beta] = - \delta_{i \alpha} \Omega^j_\beta. \ee 
The def\/ining relations on $gl(V)$, \setcounter{equation}{4}
\renewcommand{\theequation}{\arabic{section}.\arabic{equation}}
\be
[e_{i \alpha}, e_{j \beta}] = \delta_{j \alpha} e_{i \beta} - \delta_{i \beta} e_{j \alpha},
\ee and the remaining unchanged relations (7.16)--(7.18), (7.22), (7.23),
complete the picture.  We don't have to bother with the dual representation, since it's accounted for simply by replacing in the background picture each monomial $p_\alpha x_\beta$ by the monomial $-x_\alpha p_\beta$, the result of the canonical transformation\be
p_\alpha \mapsto - x_\alpha, \qquad x_\alpha \mapsto p_\alpha. \ee

However, the Lie algebra $gl(V)$ is special in that it is a Lie algebra generated by the associative algebra $End(V)$.  So, in this case,
$U({\cal G}) \approx {\cal G}$ as a vector space.  Accordingly, we can break up the dif\/ferential of the commutation relations (9.4) either as\be
[e_{i \alpha}, de_{j \beta}] = \delta_{j \alpha} de_{i \beta}
\ee or as\be
[e_{i \alpha}, d e_{j \beta}] = - \delta_{i \beta} d e_{j \alpha}.
\ee Obviously, each one of these formulae agrees with the dif\/ferential $d$ applied to the commutation relations (9.4).  It's easy to see  that formulae(9.6) and (9.7) correspond to the representations $\hat L$ and $-
\hat R$ of the Lie algebra $Lie(R)$ of an associative ring $R$
($=End(V)$ in our case) on the ring $R$ itself.  Here\setcounter{equation}{7}
\renewcommand{\theequation}{\arabic{section}.\arabic{equation}{\rm a}}
\be
\hat L_X (r) = Xr, \qquad X \in Lie (R), \quad r \in R,
\ee 
\setcounter{equation}{7}
\renewcommand{\theequation}{\arabic{section}.\arabic{equation}{\rm b}}\be
\hat R_X (r) = r X,
\ee are the left and the right multiplication operators.

Alternatively, we can treat the commutation relations (9.6) and (9.7)
in the spirit of the Diamond Lemma [1], as the rules
allowing us to move the dif\/ferentials $de_{j \beta}$'s to the left,
say, from the elements $e_{i \alpha}$'s:\setcounter{equation}{8}
\renewcommand{\theequation}{\arabic{section}.\arabic{equation}}
\be
e_{i \alpha} M_{j \beta} = M_{j \beta} e_{i \alpha} + \delta_{j \alpha} M_{i \beta},
\ee \be
e_{i \alpha} M_{j \beta} = M_{j \beta} e_{i \alpha} - \delta_{i \beta} M_{j \alpha},
\ee where $M_{j \beta}$ stands temporarily instead of the more cumbersome $d e_{j \beta}$; formulae (9.9) and (9.10) are reformulations of formulae (9.6) and (9.7) respectively.  To see  that the moving rules (9.9) and (9.10) are consistent with the commutators (9.4), we calculate, f\/irst, for the rule (9.9):\setcounter{equation}{10}
\renewcommand{\theequation}{\arabic{section}.\arabic{equation}{\rm a}}
\be
\ba{l}
e_{i \alpha} e_{j \beta} M_{k \gamma} = e_{i \alpha} (M_{k \gamma} e_{j \beta} + \delta_{k \beta} M_{j \gamma})  \\[2mm]
\ds \qquad =(M_{k \gamma} e_{i \alpha} + \delta_{k \alpha} M_{i \gamma}) e_{j \beta} + \delta_{k \beta} (M_{j \gamma} e_{i \alpha} + \delta_{j
\alpha} M_{i \gamma}).
\ea
\ee Interchanging $e_{i \alpha}$ and $e_{j \beta}$, we get
\setcounter{equation}{10}
\renewcommand{\theequation}{\arabic{section}.\arabic{equation}{\rm b}}\be
e_{j \beta} e_{i \alpha} M_{k \gamma} = (M_{k \gamma} e_{j \beta} + \delta_{k \beta} M_{j \gamma}) e_{i \alpha} + \delta_{k \alpha} (M_{i \gamma} e_{j \beta} + \delta_{i \beta} M_{j \gamma}).
\ee Subtracting (9.11b) from (9.11a), we f\/ind\setcounter{equation}{11}
\renewcommand{\theequation}{\arabic{section}.\arabic{equation}}
\be
(e_{i \alpha} e_{j \beta} - e_{j \beta} e_{i \alpha}) M_{k \gamma} = M_{k \gamma} (e_{i \alpha} e_{j \beta} - e_{j \beta} e_{i \alpha}) + \delta_{k \beta} \delta_{j \alpha} M_{i \gamma} - \delta_{k \alpha} \delta_{i \beta} M_{j \gamma}.
\ee On the other hand,\be
\ba{l}
[e_{i \alpha}, e_{j \beta}] M_{k \gamma} =  (\delta_{j \alpha} e_{i \beta} - \delta_{i \beta} e_{j \alpha}) M_{k \gamma} \\[2mm]\ds \qquad = \delta_{j \alpha} (M_{k \gamma} e_{i \beta} + \delta_{k \beta} M_{i \gamma} ) - \delta_{i \beta} (M_{k \gamma} e_{j \alpha} + \delta_{k \alpha} M_{j \gamma}),
\ea
\ee and this is the same as formula (9.12).

The relation (5.10) can be handled in the same way:
\setcounter{equation}{13}
\renewcommand{\theequation}{\arabic{section}.\arabic{equation}{\rm a}}\be
\ba{l}
e_{i \alpha} e_{j \beta} M_{k \gamma} = e_{i \alpha} (M_{k \gamma} e_{j \beta} - \delta_{j \gamma} M_{k \beta}) \\[2mm]\ds \qquad = (M_{k \gamma} e_{i \alpha} - \delta_{i \gamma} M_{k
\alpha}) e_{j \beta} - \delta_{j \gamma} (M_{k \beta} e_{i \alpha} -
\delta_{i \beta} M_{k \alpha}) \qquad  \Rightarrow
\ea
\ee \setcounter{equation}{13}
\renewcommand{\theequation}{\arabic{section}.\arabic{equation}{\rm b}}
\be
e_{j \beta} e_{i \alpha} M_{k \gamma} = (M_{k \gamma} e_{j \beta} - \delta_{j \gamma} M_{k \beta}) e_{i \alpha} - \delta_{i \gamma} (M_{k \alpha} e_{j \beta} - \delta_{j \alpha} M_{k \beta}) \qquad \Rightarrow
\ee 
\setcounter{equation}{14}
\renewcommand{\theequation}{\arabic{section}.\arabic{equation}}
\be
(e_{i \alpha} e_{i \beta} - e_{j \beta} e_{i \alpha})M_{k \gamma} = M_{k \gamma} (e_{i \alpha} e_{j \beta} - e_{j \beta} e_{i \alpha}) + \delta_{j \gamma} \delta_{i \beta} M_{k \alpha} - \delta_{i \gamma}  \delta_{j \alpha} M_{k \beta},
\ee \be
(\delta_{j \alpha} e_{i \beta} - \delta_{i \beta} e_{j \alpha}) M_{k \gamma} = M_{k \gamma} (\delta_{j \alpha} e_{i \beta} - \delta_{i \beta} e_{j \alpha}) - \delta_{j \alpha} \delta_{i \gamma} M_{k \beta}+ \delta_{i \beta} \delta_{j \gamma} M_{k \alpha},
\ee and the last two formulae are identical.
To complete our dif\/ferential complex, we have to apply the dif\/ferential $d$ to the relations (9.6) or (9.7). In each of these two cases the result is the same:\be
[d e_{i \alpha}, d e_{j \beta}]_+ = 0.
\ee Thus, we have two dif\/ferential complexes on $gl(V)$: (9.4), (9.6), (9.17) and
(9.4), (9.7), (9.17).  It would be interesting to calculate the
corresponding cohomologies.
\medskip

\noindent
{\bf Remark 9.18.} The Lie algebra $gl(V)$ has the Cartan involution $\theta$, \setcounter{equation}{17}
\renewcommand{\theequation}{\arabic{section}.\arabic{equation}{\rm a}}
\be
\theta (g) = - g^t, \qquad \theta (e_{i \alpha}) = - e_{\alpha i}. \ee 
Extended naturally to the dif\/ferentials, \setcounter{equation}{17}
\renewcommand{\theequation}{\arabic{section}.\arabic{equation}{\rm b}}
\be
\theta (de_{i \beta}) = - d e_{\beta j},
\ee the isomorphism $\theta$ interchanges the commutation rules (9.6) and (9.7).  Therefore, the two dif\/ferential complexes on $gl(V)$ are
isomorphic.
\medskip

\noindent
{\bf Remark 9.19.} Neither of the formulae (9.6), (9.7) would allow the reduction from $gl$ to $sl$.  (See  also Remark 10.14.)
\medskip

\noindent
{\bf Remark 9.20.} Each one of the two dif\/ferential complexes constructed above on the Lie algebra $gl(V)$ can be further reduced onto 4 subalgebras:  upper-triangular; lower-triangular; upper-nilpotent (upper-triangular with zeroes on the diagonal); lower-nilpotent
(lower-triangular with zeroes on the diagonal).  Some of these
subalgebras appear useful in many dif\/ferent curcumstances.  For
example, the Lie algebra $\Delta^+$ of upper-triangular matrices
underwrites the 1$^{st}$ Hamiltonian structure of the lattice KP
hierarchy
[8].  Moreover, since that
hierarchy is universal w.r.t. to its f\/inite-components cut-outs, one
immediately sees that cutting-of\/f in $\Delta^+$ all diagonals above a
f\/ixed one results in a Lie algebra as nice, w.r.t. to the
dif\/ferential-forms complex, as $\Delta^+$ itself.  Moreover still,
since the KP hierarchy can be considered either on inf\/inite or
periodic lattice, the same conclusion applies to $\Delta^+$ and all
its cut-of\/fs.  Similar considerations are pertinent for the other 3
Lie subalgebras of this Remark.
\medskip

Notice that the Lie subalgebra $\Delta^+$ provides {\it another}
generalization of the dif\/ferential complex constructed in the preceding Section for the Lie algebra $af\!f(1)$.  How dif\/ferent is it?  Consider, in each of the 2 complexes, a subcomplex generated by\setcounter{equation}{20}
\renewcommand{\theequation}{\arabic{section}.\arabic{equation}}
\be
x = e_{11}, \qquad y = e_{12}, \qquad dx = de_{11}, \qquad dy =
de_{12}.
\ee In the 1$^{st}$ complex (9.6), we have\be
[x, y] = y,
\ee \setcounter{equation}{22}
\renewcommand{\theequation}{\arabic{section}.\arabic{equation}{\rm a}}
\be
[x, dx] = dx, \qquad  [x, dy] = dy,
\ee \setcounter{equation}{22}
\renewcommand{\theequation}{\arabic{section}.\arabic{equation}{\rm b}}
\be
[y, dx] = [y, dy] = 0,
\ee 
\setcounter{equation}{22}
\renewcommand{\theequation}{\arabic{section}.\arabic{equation}{\rm c}}\be
(dx)^2 = (dy)^2 = dx dy + dy dx = 0.
\ee In the 2$^{nd}$ complex (9.7) we have\setcounter{equation}{23}
\renewcommand{\theequation}{\arabic{section}.\arabic{equation}{\rm a}}
\be
[dx, x] = dx, \qquad [dx, y] = dy,
\ee \setcounter{equation}{23}
\renewcommand{\theequation}{\arabic{section}.\arabic{equation}{\rm b}}
\be
[dy, x] = 0, \qquad [dy, y] = 0,
\ee \setcounter{equation}{23}
\renewcommand{\theequation}{\arabic{section}.\arabic{equation}{\rm c}}
\be
(dx)^2 = (dy)^2 = dx dy + dy dx = 0.
\ee Since in each case, (9.23) or (9.24), $[x, dx] \not= 0,$ we got
something quite dif\/ferent from the formulae in \S~8.  Let us look
more closely at the new formulae.
Starting with the 1$^{st}$ complex (9.22)--(9.23), let us agree  to
write elements of $U({\cal G})$, ${\cal G} = af\!f(1)$, in the normal
form
\setcounter{equation}{24}
\renewcommand{\theequation}{\arabic{section}.\arabic{equation}{\rm a}}
\be
ry^{\ldots} x^{\ldots},
\ee and elements of $\Omega^1$ in the form\setcounter{equation}{24}
\renewcommand{\theequation}{\arabic{section}.\arabic{equation}{\rm b}}
\be
dy a + dx b,
\ee with $a, b \in U ({\cal G})$ written in the normal form (9.25a).   Since, as can be easily seen by induction,\setcounter{equation}{25}
\renewcommand{\theequation}{\arabic{section}.\arabic{equation}}
\be
d(f(x)) = dx (f(x + 1) - f(x)),
\ee we arrive at the following conclusion: identifying $U({\cal G})$ with
$R[x,y]$ via formula (9.25a), the dif\/ferential $d$ on $U({\cal G}$
acts by the rule\be
d(f(y,x)) = dy f_y + dx (f(y, x+1) - f(y, x)), \qquad f \in R[x,y]
\ee 
It follows that every closed 1-form is exact.  Indeed, let the form\be
\omega = dy a + dx b, \qquad  a, b \in R[x,y],
\ee be closed:\be
a (y, x+1) - a(y, x) = b_y (y, x).
\ee Find $F \in R[x,y]$ such that\be
a = F_y,
\ee and set \be
G = b -(F(y,x + 1) - F(y, x)).
\ee Then the closedness condition (9.29) becomes\be
G_y = 0 \quad  \Rightarrow \quad G = G(x),
\ee 
and thus\be
\omega = dy F_y + dx (F(y, x+1) - F(y,x)) + dx G(x) = d(F) + dx G(x).
\ee Since the map\be
R[x]  \ni G \mapsto G (x+1) - G(x) \in R[x]
\ee is an epimorphism,\be
\{ \mbox{every  1-form} \quad  G(x) dx \quad  \mbox{is  exact} \}. \ee 
Thus, the 1-form $\omega$ (9.33) is exact as well.

In the second complex, (9.22), (9.24), the situation is similar.
Taking the normal form in $U({\cal G})$ to be \be
rx^{\ldots} y^{\ldots},
\ee and writing elements of $\Omega^1$ as\be
\omega = d x a + dy b, \qquad a, b \in R[x,y],
\ee we see  that\be
d (f(x,y)) = dx (f (x,y) - f(x -1,y)) + dy f_y.
\ee Thus, if the 1-form $\omega$ (9.37) is closed, \be
a_y = b (x, y) - b (x-1, y),
\ee we f\/irst f\/ind $F \in R[x,y]$ such that\be
b = F_y,
\ee and set\be
G = a - (F(x,y) - F(x-1,y)).
\ee 
The closedness condition (9.39) then becomes\[
G_y = 0 \qquad \Rightarrow \qquad G = G (x),
\]and thus\be
\omega = dx (F(x,y) - F(x-1,y)) + dy F_y + dx G(x) = d (F) + dx G (x), \ee 
and, again, since the map\be
R[x] \ni G (x) \mapsto G(x) - G (x-1) \in R [x]
\ee is onto, the closed 1-form $\omega$ is exact.
Thus, we have 3 dif\/ferent complexes for the Lie algebra ${\cal G} = af\!f(1)$, all with identically trivial cohomologies.

\setcounter{section}{10}
\setcounter{equation}{0}
\renewcommand{\theequation}{\arabic{section}.\arabic{equation}}

\subsection*{\S~10.  The Lie algebra {\itshape so}({\itshape n})}
Consider the Lie subalgebra $so(V)$ of $gl(V)$ consisting of skewsymmetric matrices (recall that we have f\/ixed a basis on $V$), with the basis\be
M_{ij} = - M_{ji} = e_{ij} - e_{ji}, \qquad i \not= j.
\ee The commutation relations (9.4) for $gl(V)$ imply the following
commutation relations for $so(V)$:\be
[M_{i \alpha}, M_{j \beta}] = \delta_{j \alpha} M_{i \beta} - \delta_{i \beta} M_{j \alpha} - \delta_{\alpha \beta} M_{ij} - \delta_{ij} M_{\alpha \beta} ;
\ee it is understood that $M_{ij}$ vanishes whenever $i=j$.  It's easy to see  that neither of the special $gl(V)$ relations, (9.6) or (9.7), reduces onto $so(V)$.  Hence, we have to start from scratch.

The Quantum Clebsch representation for $gl(V)$, (6.10), (9.2), \be
e_{i \alpha} = h^{-1} p_\alpha x_i,
\ee induces the corresponding representation on $so(V)$:\be
M_{i \alpha} = -h^{-1} (p_i x_\alpha - p_\alpha x_i).
\ee Hence,\be
d (M_{i \alpha}) = h^{-1} ((d p_\alpha \cdot x_i - dx_\alpha \cdot p_i) - (dp_i \cdot x_\alpha - dx_i \cdot p_\alpha)) = \theta_{i \alpha} - \theta_{\alpha i},
\ee where \be
\theta_{i \alpha} =  h^{-1} (p_\alpha dx_i - x_\alpha d  p_i).
\ee Let us next determine the commutation relations between the $M_{i \alpha}$'s and the $\theta_{j \beta}$'s.  By formulae (10.4) and (10.6),\[
\ba{l}
[M_{i \alpha}, \theta_{j \beta}] = (-h)^{-2} [p_i x_\alpha - p_\alpha x_i, x_\beta dp_j - p_\beta dx_j ] \\[2mm]
\ds \qquad =h^{-2} dp_j (x_\alpha h \delta_{i \beta} - x_i h
\delta_{\alpha \beta} ) + h^{-2} dx_j (p_j h \delta_{\beta \alpha} -
p_\alpha h \delta_{\beta i}) \\[2mm]\ds \qquad = h^{-1} \delta_{\alpha \beta} (p_i dx_j - x_i dp_j) -
h^{-1} \delta_{i \beta} (p_\alpha dx_j - x_\alpha dp_j) \
{\mathop{=}\limits^{\mbox{\scriptsize \rm  [by (10.6)]}}}\ \delta_{\alpha \beta} \theta_{j i } - \delta_{i \beta} \theta_{j
\alpha}:
\ea
\]
\be
[M_{i \alpha}, \theta_{j \beta}] = \delta_{\alpha \beta} \theta_{j i} - \delta_{i \beta} \theta_{j \alpha}.
\ee 
These  have been suggestive background calculations.  We now have to check the consistency of formulae (10.2), (10.5), (10.7).  Applying the
dif\/ferential $d$ to the LHS of formula (10.2), we get\[
\ba{l}
d ([M_{i \alpha}, M_{j \beta}]) = [M_{i \alpha}, d (M_{j \beta})] - [M_{j \beta}, d (M_{i \alpha})] \\[2mm]
\ds \qquad {\mathop{=}\limits^{\mbox{\scriptsize \rm [by (10.5)]}}} \  [M_{i \alpha}, \theta_{j \beta} - \theta_{\beta j}] - [M_{j \beta}, \theta_{i \alpha} - \theta_{\alpha i}]
\ {\mathop{=}\limits^{\mbox{\scriptsize \rm [by (10.7)]}}} \ (\delta_{\alpha \beta} \theta_{ji} - \delta_{i \beta} \theta_{j \alpha})
\\[2mm]
\ds \qquad  - (\delta_{\alpha j} \theta_{\beta i} - \delta_{ij} \theta_{\beta \alpha}) - (\delta_{\beta \alpha} \theta_{ij} - \delta_{j \alpha} \theta_{i \beta}) + (\delta_{\beta i} \theta_{\alpha j} - \delta_{j i} \theta_{\alpha
\beta} ) \\[2mm]\ds \qquad = \delta_{j \alpha} (\theta_{i \beta} - \theta_{\beta i})
- \delta_{i \beta} (\theta_{j \alpha} - \theta_{\alpha j}) - \delta_{\alpha \beta}
(\theta_{ij} -  \theta_{ji}) -  \delta_{ij} (\theta_{\alpha \beta} - \theta_{\beta \alpha}) \\[2mm]
\ds \qquad {\mathop{=}\limits^{\mbox{\scriptsize \rm [by (10.5)]}}} \ \delta_{j \alpha} d(M_{i \beta}) - \delta_{i \beta} d (M_{j \alpha} ) - \delta_{\alpha \beta} d (M_{ij}) - \delta_{ij} d (M_{\alpha
\beta}),
\ea
\]and this is the dif\/ferential of the $RHS$ of formula (10.2).
Now, set\be
d (\theta_{i \alpha}) = \rho_{i \alpha} = \rho_{\alpha i},
\ee \be
d (\rho_{i \alpha}) = 0.
\ee By formula (10.6),\be
\rho_{i \alpha} = d (\theta_{i \alpha}) = h^{-1} (d p_\alpha dx_i -
dx_\alpha d p_i);
\ee we thus impose the commutation relations\setcounter{equation}{10}
\renewcommand{\theequation}{\arabic{section}.\arabic{equation}{\rm a}}
\be
[\rho_{i \alpha}, M_{j \beta}] = 0,
\ee 
\setcounter{equation}{10}
\renewcommand{\theequation}{\arabic{section}.\arabic{equation}{\rm b}}\be
[\rho_{i \alpha}, \theta_{j \beta}] = 0,
\ee 
\setcounter{equation}{10}
\renewcommand{\theequation}{\arabic{section}.\arabic{equation}{\rm c}}\be
[\rho_{i \alpha}, \rho_{j \beta}] = 0.
\ee Applying the dif\/ferential $d$ to the relations (10.7), we get
\setcounter{equation}{11}
\renewcommand{\theequation}{\arabic{section}.\arabic{equation}}
\be
[\theta_{i \alpha} - \theta_{\alpha i}, \theta_{j \beta}]_+ + [M_{i
\alpha}, \rho_{j \beta}] = \delta_{\alpha \beta} \rho_{j i} -
\delta_{i \beta} \rho_{j \alpha} .
\ee Thus, we nee d to determine $[\theta_{i \alpha}, \theta_{j \beta}]_+$'s.  Using the background formulae (10.6), we f\/ind\[
\ba{l}
\hspace*{-11.8pt}[\theta_{i \alpha}, \theta_{j \beta}]_+ = h^{-2} ((p_\alpha dx_i - x_\alpha d p_i) (p_\beta dx_j - x_\beta dp_j) + (p_\beta dx_j - x_\beta dp_j) (\rho_\alpha dx_i - x_\alpha d p_i)) \\[1mm]\ds \qquad  = - h^{-2} (x_\alpha p_\beta d p_i dx_j + p_\alpha x_\beta dx_i dp_j + x_\beta p_\alpha dp_j dx_j + p_\beta x_\alpha dx_j d p_i)\\[1mm]
\ds \qquad = h^{-2} dp_i dx_j (- x_\alpha p_\beta + p_\beta x_\alpha)
- h^{-2} dx_i d p_j (p_\alpha x_\beta - x_\beta p_\alpha) \\[1mm]
\ds \qquad = \delta_{\alpha \beta} h^{-1} (d p_j dx_j - dx_i dp_j) \
{\mathop{=}\limits^{\mbox{\scriptsize \rm [by (10.10)]}}} \
\delta_{\alpha \beta} \rho_{ji};
\ea
\]
\be
[\theta_{i \alpha}, \theta_{j \beta}]_+ = \delta_{\alpha \beta} \rho_{ji}.
\ee Taking this formula as a new relation, substituting it into the $LHS$ of formula (10.12), and remembering formula (10.11a), we f\/ind\[
[\theta_{i \alpha}, \theta_{j \beta}]_+ - [\theta_{\alpha i}, \theta_{j \beta}]_+ = \delta_{\alpha \beta} \rho_{j i} - \delta_{i \beta} \rho_{j \alpha},
\]and this is the $RHS$ of formula (10.12).  It remains to apply the
dif\/ferential $d$ to each of the relations (10.11a,b,c), (10.13), and in each case
we get an identitically satisf\/ied relation.

Thus, the dif\/ferential complex $\Omega^*$ on $so(V)$ has:   1) the generators $\{M_{ij} = - M_{ji}, i \not= j\},$ $\{\theta_{ij}\}$, $\{\rho_{ij} = \rho_{ji}\}$; 2) the action of the dif\/ferential $d$,
(10.5), (10.8) (10.9); 3) and the relations (10.2), (10.7), (10.11), (10.13).  We see  that
in addition to the generators $M_{ij}$'s of ${\cal G}$, we had to introduce
some extra generators, $\theta_{ij}$'s and $\rho_{ij}$'s, to complete the complex $\Omega^*
U(so(n))$; however, the number of extra generators has turned out to
be smaller than what one would have expected from the general
formulae of \S~7.
\setcounter{equation}{14}

\medskip

\noindent
{\bf Remark 10.14}.  Is it possible to construct a dif\/ferential forms complex on the Lie algebra $so(n)$ (or other semi-simple Lie algebras) without introducing Quantum ghosts?  It seems unlikely.  Let us look, for example, at the f\/irst nontrival case, ${\cal G} = so(3) \approx
sl(2).$ According to formulae (7.4), (7.5), we need to chose a
3-dimensional representation of ${\cal G}$.  So, it's either the
direct sum of 1-dimensional trivial and 2-dimensional fundamental, or
the adjoint representation.  The 1$^{st}$ alternative can be ruled out
in view of the 1-dimensional representation being trivial.  This
leaves the adjoint representation.  In the standard basis $e,f,h$ of
$sl(2)$, we thus must have
\be
\left(\matrix{de \cr df \cr dh \cr} \right)
= {\cal{M}} \left(\matrix{e \cr f \cr h \cr} \right),
\ee with some constant nondegenerate matrix ${\cal{M}}$; this formulae is
to be understood not literary, but only as describing the action of
${\cal G}$ on $d({\cal G})$.  Now, the consistency conditions (7.4)
imply that the matrix ${\cal{M}}$  has the form\be
{\cal{M}} = \left(\matrix{\lambda & 0 & - \nu \cr0 & - \lambda & \mu \cr -2 \mu & 2 \nu & 0 \cr} \right).
\ee But $\det ({\cal{M}}) = 0$, so the ghostless complex of dif\/ferential
forms on ${\cal G} = sl (2)$ doesn't exist.

\setcounter{section}{11}
\setcounter{equation}{0}
\renewcommand{\theequation}{\arabic{section}.\arabic{equation}}

\subsection*{\S~11.  {\itshape Q}-Quantum spaces}
In the associative ring $R\langle x\rangle$, consider the commutation
relations
\be
x_i x_j = Q_{ij} x_j x_i, \qquad \forall \ i, j,
\ee where $Q_{ij}$'s are arbitrary invertible constants,\be
Q_{ij} = Q_{ji}^{-1}, \qquad Q_{ii} = 1;
\ee if the $Q_{ij}$'s do not belong initially to the ring $R$, we can always adjoin them.
Let us construct a complex of dif\/ferential forms over our ring
$R_Q\langle x\rangle$.  To do that, we need to postulate the
commutation relations between $x_i$'s and $dx_j$'s.  From the
experience of Quantum Groups, one knows that there is no canonical way
to extend relations from a ring into the corresponding
dif\/ferential-forms ring; such extensions may vary with the situation
at hand and with the imagination of the extender.  With this in mind,
let us proceed in the engineering spirit of this paper, taking the
view that $dx_i$ is ``a very small increment in the variable
$x_i$'', and thus $dx_i$ should have the same commutation relations as $x_i$ does, to wit:\be
(dx_i) x_j = Q_{ij} x_j dx_i, \qquad \forall \ i,j.
\ee In particular, \be
[dx_i, x_i] = 0, \qquad \forall \ i.
\ee 
Before proceeding further, we have to verify that the commutation rules (11.1) and (11.3) are compatible.  Applying the dif\/ferential $d$ to the relation (11.1) and keeping in mind that $d$ is a derivation, we f\/ind\[
\ba{l}
d (x_i x_j - Q_{ij} x_j x_i) = (dx_i) x_j + x_i dx_j - Q_{ij} (dx_j) x_j - Q_{ij} x_j dx_i \\[2mm]
\ds \qquad = ((dx_i) x_j - Q_{ij} x_j dx_i) - Q_{ij} ((dx_j) x_i - Q_{ji} x_i dx_j),
\ea
\]and each of these 2 summands vanishes by formulae (11.3).
Finally, applying the dif\/ferential $d$ to the relation (11.3) and remembering that $d$ is a ${\bf Z}_2$-graded derivations, we get\be
dx_i dx_j = - Q_{ij} dx_j dx_i, \qquad \forall \ x,j.
\ee 
The dif\/ferential complex $\{\Omega^*, d\}$ results thereby.  (In this and subsequent Sections, all the variables are considered bosonic.  A more general case, on the lines of \S~4, is left  to the reader.)
Let us ascertain whether the cohomologies of our complex are trivial
or not.
Proceeding as in \S~2, we extend $R_Q\langle x\rangle $ and
$\Omega^*$ by adjoining a new variable $t$ {\it commuting} {\it with  everything}, with its dif\/ferential $\tau = dt$ behaving accordingly, i.e., commuting with everything in the ${\bf Z}_2$-graded sense.  Denoting the extended dif\/ferential-forms ring by $\bar \Omega^*$, we again have:  the unique decomposition\be
\omega = \omega_+ + \tau \omega_-, \qquad \forall \ \omega \in \bar
\Omega^*,
\qquad \omega_\pm \in \Omega^* [t];
\ee the homotopy operator\be
I: \ \bar \Omega^* \rightarrow \Omega^*,
\ee \be
I (\omega) = \int\limits^1_0 dt\, \omega_-;
\ee and the ring homomorphism $A_t: \Omega^* \rightarrow \bar \Omega^* $ (over $R$), \setcounter{equation}{8}
\renewcommand{\theequation}{\arabic{section}.\arabic{equation}{\rm a}}
\be
A_t (x_i) = t x_i, \qquad \forall \ i,
\ee 
\setcounter{equation}{8}
\renewcommand{\theequation}{\arabic{section}.\arabic{equation}{\rm b}}\be
A_t (dx_i) = t dx_i + \tau x_i, \qquad \forall \ i,
\ee so that\setcounter{equation}{9}
\renewcommand{\theequation}{\arabic{section}.\arabic{equation}}
\be
A_t d = d A_t.
\ee To make sure that the homomorphism $A_t$ is well-def\/ined, we have to verify that the relations (11.1), (11.3), (11.5) are preserved when acted
upon by $A_t$.  So, \setcounter{equation}{10}
\renewcommand{\theequation}{\arabic{section}.\arabic{equation}{\rm a}}
\be
A_t (x_i x_j - Q_{ij} x_j x_i) = t^2 (x_i x_j - Q_{ij} x_j x_i),
\ee 
\setcounter{equation}{10}
\renewcommand{\theequation}{\arabic{section}.\arabic{equation}{\rm b}}
\be
\ba{l}
\ds A_t ((dx_i) x_i - Q_{ij} x_j dx_i) = (t dx_i + \tau x_i) tx_j - Q_{ij} tx_j (t dx_i + \tau x_i) \\[2mm]
\ds \qquad =t^2 ((dx_i ) x_j - Q_{ij} x_j dx_i) + t \tau( x_i x_j - Q_{ij} x_j x_i),
\ea
\ee \setcounter{equation}{10}
\renewcommand{\theequation}{\arabic{section}.\arabic{equation}{\rm c}}
\be
\ba{l}
A_t (dx_i dx_j + Q_{ij} dx_j dx_i) = (t dx_i + \tau x_i) (t dx_j + \tau x_j) \\[2mm]
\ds \qquad + Q_{ij} (t dx_j + \tau x_j) (t dx_i + \tau x_i)  =t^2 (dx_i dx_j + Q_{ij} dx_j dx_i) \\[2mm]
\ds \qquad - t\tau  \langle ((dx_i ) x_j - Q_{ij} x_j dx_i) + Q_{ij} ((dx_j) x_i - Q_{ji} x_i dx_j ) \rangle,
\ea
\ee and we can now proceed to establish the homotopy formula:
\setcounter{equation}{12}
\renewcommand{\theequation}{\arabic{section}.\arabic{equation}}

\medskip

\noindent
{\bf Lemma 11.12.} {\it  For any $\omega \in \bar \Omega^*,$ \be
I d(\omega) + d I (\omega) = \omega_+|_{t=1} - \omega_+|_{t=0},
\ee 
}
\noindent
{\bf Proof.}  It's enough to consider two separate cases:  $\omega = t^n \nu$ and $\omega = \tau t^n \nu$, $\nu \in \Omega^*$, $n \in {\bf Z}_+$.

(A)  If $\omega = t^n \nu$ then $\omega_- = 0$, so that $I (\omega) = 0$, and hence\[
Id (\omega) = Id (t^n \nu) = I \left(n t^{n-1} \tau \nu \right) =
\int\limits^1_0 n t^{n-1} dt \,\nu = \nu \left(1 - \delta^0_n\right) = t^n \nu|_{t=1} - t^n
\nu|_{t=0};
\]

(B)  If $\omega = \tau t^n$, then $\omega_+ = 0$, and \[
\hspace*{-9.8pt}
Id (\omega) + d I(\omega) = I \left(- \tau t^n d (\nu) \right) + d
\left( \int\limits^1_0  t^n dt \,\nu\right) =
- \int\limits^1_0 t^n dt \,d(\nu) + \int\limits^1_0 t^n dt \,d(\nu) = 0. \quad
\mbox{\rule{2mm}{4mm}}\]

\setcounter{equation}{14}

\noindent
{\bf Corollary 11.14.} {\it  Every closed form $\omega \in \Omega^*$ differs from an exact one by an element from $R$.}

\medskip
\noindent
{\bf Proof.} If $\omega$ is closed, $d(\omega) = 0,$ then so is $A_t (\omega)$.  Therefore, by formula (11.13) applied to $A_t
(\omega)$, \be
d I A_t (\omega) = \omega - p r^{0,0} (\omega). \qquad
\mbox{\rule{2mm}{4mm}}\ee 

So far we have treated dif\/ferential forms as self-important
entities, without any reference to vector f\/ields.  The reason for
this reticience is a common bane of Quantum mathematics:  there exist
very few vector f\/ields, and whenever they do exist, their values on
the generators $x_i$'s are far from arbitrary.  It's easy to
understand why this is so:  any Quantum derivation has to preserve all
the def\/ining commutation relations (11.1) (or similar ones in more
general Quantum circumstances), and this is,

in general, close to impossible.  This is the chief reason the
traditional approach to the variational calculus, either commutative
[10]
or noncommutative one [12], has to be abandoned in the
Quantum framework. But some useful things can be salvaged.

Among the latter are (left) partial derivatives $\ds {\partial \over \partial x_k}$'s.  They are {\it not} derivatives any more, but are instead additive maps over $R$, satisfying the properties\be
{\partial \over \partial x_k} (r) = 0, \qquad {\partial \over \partial x_k} r = r {\partial \over \partial x_k},  \qquad \forall \ r \in R,
\ee \be
{\partial \over \partial x_k} x_i = \delta_{ik} + Q_{ik} x_i {\partial \over \partial x_k}, \qquad \forall \ i, k.
\ee Denote\be
P_{ij} = x_i x_j - Q_{ij} x_j x_i.
\ee By formula (11.17), we have\be
\ba{l}
\ds {\partial \over \partial x_k} x_i x_j =
\left(\delta_{ik} + Q_{ik} x_i {\partial \over \partial x_k}\right) x_j = \delta_{ik} x_j + Q_{ik} x_i \left(\delta_{kj} + Q_{jk} x_j {\partial \over \partial
x_k}\right) \\[3mm]
\ds \qquad = \delta_{ik} x_j + \delta_{jk} Q_{ik} x_i + Q_{ik} Q_{jk} x_i x_j {\partial \over \partial x_k}.
\ea
\ee Therefore,\be
\ba{l}
\ds {\partial \over \partial x_k} P_{ij} = \left(\delta_{ik} x_j +
\delta_{jk} Q_{ik} x_i + Q_{ik} Q_{jk} x_i x_j {\partial \over \partial x_k}\right) \\[4mm]
\ds \qquad - Q_{ij} \left(\delta_{jk} x_i + \delta_{ik} Q_{jk} x_j +
Q_{jk} Q_{ik} x_j x_i {\partial \over \partial x_k}\right) =
 Q_{ik} Q_{jk} P_{ij} {\partial \over \partial x_k} \\[3mm]
\ds \qquad + \delta_{ik} x_j (1 - Q_{jk} Q_{ij}) + \delta_{jk} x_i
(Q_{ik} - Q_{ij}) = Q_{ik} Q_{jk} P_{ij} {\partial \over \partial x_k}. \ea
\ee 
Thus, the partial derivatives $\ds {\partial \over \partial x_k}$'s are well-def\/ined.  Their connection with dif\/ferential forms is described by the following
\setcounter{equation}{21}

\medskip

\noindent
{\bf Lemma 11.21.} {\it Denote by\be
X = \sum_k dx_k {\partial \over \partial x_k}
\ee the additive map (over $R$) from $R_Q \langle x\rangle $ into
$\Omega^1$.  (The sum is well-defined even if the number of
generators $x_k$'s is infinite.)  Then:

\mbox{\rm (i)} $X$ is a derivation:\be
X (HF) = X(H)F + H X(F), \qquad \forall \ H, F \in R_Q\langle
x\rangle ;
\ee 
\mbox{\rm (ii)} $X = d$:\be
d (H) = \sum_k dx_k {\partial H \over \partial x_k}, \qquad
 \forall \ H \in R_Q \langle x\rangle .
\ee 
}
\noindent
{\bf  Proof.} (i)  We have,\be
\ba{l}
\ds X x_s = \sum dx_k {\partial \over \partial x_k} x_s \
{\mathop{=}\limits^{\mbox{\scriptsize \rm [by (11.17)]}}} \ \sum dx_k \left(\delta_{ks} + Q_{sk} x_s {\partial \over \partial
x_k} \right)\\[4mm] \ds \qquad {\mathop{=}\limits{\mbox{\scriptsize  \rm [by (11.3)]}}} \ dx_s + \sum_k x_s dx_k {\partial \over \partial x_k} = X (x_s) + x_s X.
\ea
\ee By induction on $\mbox{deg}_x (H)$, it follows that\be
X H = X (H) + HX, \qquad \forall \ H,
\ee and this is equivalent to the derivation property (i); \\(ii) Both $X$ and $d$ are derivations over $R$, sending  $x_s$ into $dx_s$ for all $s$.  Hence, $X = d$.  \hfill \rule{2mm}{4mm}
\setcounter{equation}{27}

\medskip

\noindent
{\bf Remark  11.27.} Denote by $\Omega^\ell$ the $R\langle x\rangle $-bimodule of $\ell$-forms in $\Omega^*$.  The
previous Lemma shows that instead of the general associative def\/inition\be
\Omega^1 = \left\{ \sum_{ks} f_{ks} dx_k g_{ks} \ | \ f_{ks}, g_{ks} \in R \langle x\rangle  \right\},
\ee in the $Q$-picture we can take $\Omega^1$ as\be
\Omega^1 = \left\{\sum dx_k f_k \ | \ f_k \in R_Q \langle x\rangle
\right \}.
\ee 
Similar observation applies to $\Omega^\ell$:  we can move all $\ell$ $dx$'s to the left in each monomial in a $\ell$-form $\omega \in \Omega^\ell$.
\medskip

\noindent
{\bf Remark 11.30.} The partial derivatives $\ds {\partial \over
\partial x_k}$'s are no longer derivations, as their def\/ining
formula (11.17) shows;
they should be called $Q$-derivations instead.
Nevertheless, these partial derivatives almost commute betwee n themselves:
\setcounter{equation}{31}

\medskip

\noindent
{\bf Lemma 11.31.} \be
{\partial \over \partial x_k} {\partial \over \partial x_\ell} = Q_{k\ell} {\partial \over \partial x_\ell } {\partial \over \partial x_k}, \qquad  \forall \ k, \ell.
\ee 
\noindent
{\bf Proof.} Denote\be
{\cal O}_{k \ell} = {\partial \over \partial x_k} {\partial \over \partial x_\ell} - Q_{k \ell} {\partial \over \partial x_\ell} {\partial \over
\partial x_k}.
\ee Then ${\cal O}_{k\ell}$ is an additive map over $R$ which annihilates
$R$  and the $x_s$'s.  Further, \[
\ba{l}
\ds {\cal O}_{k\ell} x_s = \left({\partial \over \partial x_k}
{\partial \over \partial x_\ell} - Q_{k\ell} {\partial \over \partial x_\ell} {\partial \over \partial x_k} \right) x_s \\[3mm]
\ds \qquad  {\mathop{=}\limits^{\mbox{\scriptsize \rm [by 11.17)]}}} \ {\partial \over \partial x_k} \circ \left( \delta_{\ell s} + Q_{s \ell}  x_s {\partial \over \partial x_\ell} \right) - Q_{k \ell} {\partial \over \partial x_\ell} \circ \left( \delta_{ks} + Q_{sk} x_s {\partial \over \partial x_k} \right) \\[3mm]
\ds \qquad = \delta_{\ell s} {\partial \over \partial x_k} + Q_{s \ell} \left(\delta_{ks} + Q_{sk} x_s {\partial \over \partial x_k} \right) {\partial \over \partial x_\ell} - Q_{k \ell} \delta_{ks} {\partial \over \partial x_\ell} \\[3mm]
\ds \qquad - Q_{k \ell} Q_{sk} \left(\delta_{\ell s} + Q_{s \ell} x_s
{\partial \over \partial x_\ell} \right) {\partial \over \partial
x_k} = \delta_{\ell s} {\partial \over \partial x_k} (1 - Q_{k \ell}
Q_{s k}) + \delta_{ks} {\partial \over \partial x_\ell} (Q_{s \ell} -
Q_{k \ell})\\[3mm]
\ds \qquad + Q_{sk} Q_{s \ell} x_s \left( {\partial \over \partial x_k} {\partial \over \partial x_\ell} - Q_{k \ell} {\partial \over
\partial x \ell} {\partial \over \partial x_k} \right) =
Q_{sk} Q_{s \ell}x_s \left( {\partial \over \partial x_k} {\partial \over \partial x_\ell} - Q_{k \ell} {\partial \over \partial x_\ell} {\partial \over \partial x_k}
\right)\!.
\ea
\]Thus, \be
{\cal O}_{k \ell} x_s = Q_{sk} Q_{s \ell} x_s {\cal O}_{k \ell}, \qquad \forall \ s.
\ee Therefore, ${\cal O}_{k \ell} = 0$. \qquad \rule{2mm}{4mm}

\setcounter{section}{12}
\setcounter{equation}{0}
\renewcommand{\theequation}{\arabic{section}.\arabic{equation}}

\subsection*{\S~12.  {\itshape Q}-Quantum spaces and discrete groups}
When one considers a discrete version of a physical or mathematical picture, the basic variables acquire discrete indices, either of a discrete group $G$ or its homogeneous space.  Most often one has ${\bf Z}$,
${\bf Z}_N$, and their products as the underlying group, but in
certain constructions it is
easier to work with an arbitrary unspecif\/ied group.  This is what we shall do in this Section.
Suppose, in the language of the preceding Section, that our variables carry two indices, $i$ and $g$:  $x_i^{(g)}$, where letters $f, g, h$ in this Section are reserved for typical elements of the f\/ixed discrete group $G$.  The group $G$ acts on $R_Q \langle x\rangle$ by automorphisms,
with the action on the generators by the rule\be
\hat h (x_i^{(g)} ) = x_i^{(hg)}, \qquad \forall \ i, \quad \forall \
h, g \in G.
\ee Further, the commutation relations between the $x_i^{(g)}$'s are assumed to be $G$-invariant:\be
x_i^{(g)} x_j^{(h)} = Q_{ij}^{g^{-1}h} x_j^{(h)} x^{(g)}_i, \qquad\forall \ i, j, \qquad \forall \ g, h \in G.
\ee 
Also, the actions of $G$ and $d$ on $\Omega^*$ commute:\be
\hat g d = d \hat g, \qquad \forall \ g \in G.
\ee 
If nothing else intervenes, the results of \S~11 remain true as there stated:  every closed $\ell$-form is exact for $\ell > 0$.  But suppose we introduce into $\Omega^*$ the equivalence relation of equivariance:\be
\omega_1 \sim \omega_2 \quad \Leftrightarrow \quad  \exists \ g \in G: \quad\hat g (\omega_1) = \omega_2.
\ee 
\setcounter{equation}{5}

\noindent
{\bf Lemma 12.5.} {\it Suppose $\omega \in \Omega^\ell, \ell > 0$, and $d (\omega) \sim 0$.  Then there exists $\nu \in \Omega^{\ell -1} $ such that $\omega \sim d (\nu)$.}

\medskip

\noindent{\bf Proof.}  We proceed as in the preceding Section, by adding one more variable $t$ on which $G$ {\it acts  trivially}: \be
\hat g (t) = t, \qquad \forall \ g \in G .
\ee Then we again get the homotopy formula\be
I d (\bar \omega) + d I (\bar \omega)= \bar \omega_+|_{t=1} - \bar
\omega _+|_{t=0}, \qquad \forall \ \bar \omega \in \bar \Omega^*.
\ee 
Taking\be
\bar \omega = A_t (\omega),
\ee and noticing that\be
A_t \hat g = \hat g A_t, \qquad \forall \ g \in G,
\ee \be
I \hat g = \hat g I, \qquad \forall \ g \in G,
\ee we f\/ind that\be
\omega = d (I A_t (\omega)) + I A_t d (\omega).
\ee Thus, if $d(\omega)$ is trivial, i.e., $d (\omega) \sim 0$, then so is $\omega - d (\nu)$, $\nu = IA_t (\omega)$. \qquad \rule{2mm}{4mm}

It is an entirely dif\/ferent matter to describe by {\it differential equations} not simply exact dif\/ferential forms, as in the
Poincar\'e Lemma, but just the trivial ones (w.r.t. the action of the
group $G$.) The machinery to perform such feats is customarily called
the {\it Variational Calculus}.  This will be developed in the
4$^{th}$ Act.

 \label{kupershmidt_1-lp}

\noindent
The * attached to a citation marks the publisher that demands, as a
condition of printing a paper in the publisher's journal, that all
authors of creative works surrender and hand in to the publisher the
copyright to the fruits of their labors. The publishers so noted do
not include those in Cuba, Iraq,
North Korea, and other savage places, where such a policy is not a
matter of free choice but is state-mandated.

\end{document}